\documentclass[pra,aps,twocolumn,twoside,superscriptaddress]{revtex4}

\usepackage{amsmath,amsfonts,amssymb,color,graphics,graphicx,latexsym,revsymb,amsthm,url,verbatim,appendix,epstopdf}
\usepackage{algorithm}

\usepackage{algcompatible}

\usepackage{hyperref}

\floatname{algorithm}{Protocol}

\newtheorem{theorem}{Theorem}
\newtheorem{corollary}[theorem]{Corollary}
\newtheorem{lemma}[theorem]{Lemma}
\newtheorem{proposition}[theorem]{Proposition}

\theoremstyle{definition}
\newtheorem{definition}{Definition}

\def\squareforqed{\hbox{\rlap{$\sqcap$}$\sqcup$}}
\def\qed{\ifmmode\squareforqed\else{\unskip\nobreak\hfil
\penalty50\hskip1em\null\nobreak\hfil\squareforqed
\parfillskip=0pt\finalhyphendemerits=0\endgraf}\fi}
\def\endenv{\ifmmode\;\else{\unskip\nobreak\hfil
\penalty50\hskip1em\null\nobreak\hfil\;
\parfillskip=0pt\finalhyphendemerits=0\endgraf}\fi}
\def\Dbar{\leavevmode\lower.6ex\hbox to 0pt
{\hskip-.23ex\accent"16\hss}D}

\makeatletter
\def\bcj{\begin{conjecture}}
\def\ecj{\end{conjecture}}
\def\bcr{\begin{corollary}}
\def\ecr{\end{corollary}}
\def\bd{\begin{definition}}
\def\ed{\end{definition}}
\def\bea{\begin{eqnarray}}
\def\eea{\end{eqnarray}}
\def\bem{\begin{enumerate}}
\def\eem{\end{enumerate}}
\def\bex{\begin{example}}
\def\eex{\end{example}}
\def\bim{\begin{itemize}}
\def\eim{\end{itemize}}
\def\bl{\begin{lemma}}
\def\el{\end{lemma}}
\def\bpf{\begin{proof}}
\def\epf{\end{proof}}
\def\bpp{\begin{proposition}}
\def\epp{\end{proposition}}
\def\bqu{\begin{question}}
\def\equ{\end{question}}
\def\br{\begin{remark}}
\def\er{\end{remark}}
\def\bt{\begin{theorem}}
\def\et{\end{theorem}}

\def\btb{\begin{tabular}}
\def\etb{\end{tabular}}

\newcommand{\nc}{\newcommand}

\def\a{\alpha}
\def\b{\beta}
\def\g{\gamma}

\def\e{\epsilon}

 \nc{\bA}{{\bf A}} \nc{\bB}{{\bf B}} \nc{\bC}{{\bf C}}
 \nc{\bD}{{\bf D}} \nc{\bE}{{\bf E}} \nc{\bF}{{\bf F}}
 \nc{\bG}{{\bf G}} \nc{\bH}{{\bf H}} \nc{\bI}{{\bf I}}
 \nc{\bJ}{{\bf J}} \nc{\bK}{{\bf K}} \nc{\bL}{{\bf L}}
 \nc{\bM}{{\bf M}} \nc{\bN}{{\bf N}} \nc{\bO}{{\bf O}}
 \nc{\bP}{{\bf P}} \nc{\bQ}{{\bf Q}} \nc{\bR}{{\bf R}}
 \nc{\bS}{{\bf S}} \nc{\bT}{{\bf T}} \nc{\bU}{{\bf U}}
 \nc{\bV}{{\bf V}} \nc{\bW}{{\bf W}} \nc{\bX}{{\bf X}}
 \nc{\bZ}{{\bf Z}}

\nc{\cA}{{\cal A}} \nc{\cB}{{\cal B}} \nc{\cC}{{\cal C}}
\nc{\cD}{{\cal D}} \nc{\cE}{{\cal E}} \nc{\cF}{{\cal F}}
\nc{\cG}{{\cal G}} \nc{\cH}{{\cal H}} \nc{\cI}{{\cal I}}
\nc{\cJ}{{\cal J}} \nc{\cK}{{\cal K}} \nc{\cL}{{\cal L}}
\nc{\cM}{{\cal M}} \nc{\cN}{{\cal N}} \nc{\cO}{{\cal O}}
\nc{\cP}{{\cal P}} \nc{\cQ}{{\cal Q}} \nc{\cR}{{\cal R}}
\nc{\cS}{{\cal S}} \nc{\cT}{{\cal T}} \nc{\cU}{{\cal U}}
\nc{\cV}{{\cal V}} \nc{\cW}{{\cal W}} \nc{\cX}{{\cal X}}
\nc{\cZ}{{\cal Z}}

\nc{\hA}{{\hat{A}}} \nc{\hB}{{\hat{B}}} \nc{\hC}{{\hat{C}}}
\nc{\hD}{{\hat{D}}} \nc{\hE}{{\hat{E}}} \nc{\hF}{{\hat{F}}}
\nc{\hG}{{\hat{G}}} \nc{\hH}{{\hat{H}}} \nc{\hI}{{\hat{I}}}
\nc{\hJ}{{\hat{J}}} \nc{\hK}{{\hat{K}}} \nc{\hL}{{\hat{L}}}
\nc{\hM}{{\hat{M}}} \nc{\hN}{{\hat{N}}} \nc{\hO}{{\hat{O}}}
\nc{\hP}{{\hat{P}}} \nc{\hR}{{\hat{R}}} \nc{\hS}{{\hat{S}}}
\nc{\hT}{{\hat{T}}} \nc{\hU}{{\hat{U}}} \nc{\hV}{{\hat{V}}}
\nc{\hW}{{\hat{W}}} \nc{\hX}{{\hat{X}}} \nc{\hZ}{{\hat{Z}}}

\nc{\hn}{{\hat{n}}}

\def\max{\mathop{\rm max}}

\def\tr{\mathop{\rm Tr}}

\def\ox{\otimes}

\newcommand{\bra}[1]{\langle#1|}
\newcommand{\ket}[1]{|#1\rangle}
\newcommand{\proj}[1]{| #1\rangle\!\langle #1 |}
\newcommand{\ketbra}[2]{|#1\rangle\!\langle#2|}
\newcommand{\braket}[2]{\langle#1|#2\rangle}

\begin{document}
\title{Check-based generation of one-time tables using qutrits}
\author{Li Yu}\email{yuli@hznu.edu.cn}
\affiliation{School of Physics, Hangzhou Normal University, Hangzhou, Zhejiang 311121, China}
\author{Xue-Tong Zhang}
\affiliation{School of Physics, Hangzhou Normal University, Hangzhou, Zhejiang 311121, China}
\author{Fuqun Wang}
\affiliation{School of Mathmatics, Hangzhou Normal University, Hangzhou, Zhejiang 311121, China}
\affiliation{Key Laboratory of Cryptography of Zhejiang Province, Hangzhou 311121, China}
\affiliation{Westone Cryptologic Research Center, Beijing 100071, China}
\author{Chui-Ping Yang}\email{yangcp@hznu.edu.cn}
\affiliation{School of Physics, Hangzhou Normal University, Hangzhou, Zhejiang 311121, China}
\affiliation{Quantum Information Research Center, Shangrao Normal University, Shangrao, Jiangxi 334001, China}

\begin{abstract}
One-time tables are a class of two-party correlations that can help achieve information-theoretically secure two-party (interactive) classical or quantum computation. In this work we propose a bipartite quantum protocol for generating a simple type of one-time tables (the correlation in the Popescu-Rohrlich nonlocal box) with partial security. We then show that by running many instances of the first protocol and performing checks on some of them, asymptotically information-theoretically secure generation of one-time tables can be achieved. The first protocol is adapted from a protocol for semi-honest quantum oblivious transfer, with some changes so that no entangled state needs to be prepared, and the communication involves only one qutrit in each direction. We show that some information tradeoffs in the first protocol are similar to that in the semi-honest oblivious transfer protocol. We also obtain two types of inequalities about guessing probabilities in some protocols for generating one-time tables, from a single inequality about guessing probabilities in semi-honest quantum oblivious transfer protocols.
\end{abstract}
\maketitle


\section{Introduction}\label{sec1}

Many problems in classical cryptography are special cases of the secure two-party function evaluation problem. The goal of such problem is to correctly compute some function of the inputs from the two parties, while keeping the inputs as private from the opposite party as possible. Possible approaches to this problem include classical homomorphic encryption \cite{Gentry09,brakerski2011efficient}, or Yao's ``Garbled Circuit'' \cite{Yao86} and its variants. Another possibility is to introduce a trusted third party, who may sometimes interact with the two parties for multiple rounds. To lower the requirement on the trusted third party, a ``trusted initializer'' has been proposed \cite{Beaver98}. Such trusted initializer only prepares some initial correlations between the two parties, and does not interact with any party afterwards. Such initial correlations are often called ``one-time tables'', and a simplest type is the correlations present in the Popescu-Rohrlich nonlocal box \cite{Popescu1994}.

Secure two-party quantum computation is the corresponding problem in quantum computing and quantum cryptography. The two parties wish to correctly compute an output according to some public or private program while keeping their (quantum) inputs as secure as possible. Special cases of this general problem include quantum homomorphic encryption (QHE) \cite{rfg12,MinL13,ypf14,Tan16,Ouyang18,bj15,Dulek16,NS17,Lai17,Mahadev17,ADSS17,Newman18,TOR18}, secure assisted quantum computation \cite{Ch05,Fisher13}, computing on shared quantum secrets \cite{Ouyang17}, and physically-motivated secure computation (e.g. \cite{OTF20}). In the study of QHE, it is found that secure computation of the modulo-$2$ inner product of two bit strings provided by the two parties is a key task, and the one-time tables mentioned above turn out to be helpful for this task.

In this work, we firstly propose a simple two-party quantum protocol, using a qutrit in two directions of communication, for generating one-time tables with partial security. It is adapted from the semi-honest oblivious transfer protocol in \cite{CKS13,CGS16} with significant changes. We then show that by allowing for checks and the associated possible aborts, such protocol can be enhanced to achieve asymptotic information-theoretic security. We provide some analysis of the tradeoff relations of mutual information, Holevo bounds, or guessing probabilities arising from the protocols. The first protocol, Protocol~\ref{ptl:NLAND}, implements the following task with partial privacy: it takes as input two locally-generated uniformly random bits $x$ and $y$ from Alice and Bob, respectively, and outputs $(x\,\rm{AND}\,y)\,\rm{XOR}\,r$ on Alice's side and $r$ on Bob's side, where $r$ is a uniformly random bit. This implies that our type of one-time table contains four bits: two input bits and two output bits.

Security in quantum key distribution \cite{BB84} is dependent on verifications. Inspired by this, we propose some protocols that verify the correctness of Protocol~\ref{ptl:NLAND}. We propose Protocol~\ref{ptl:precompute1} to select some one-time tables generated by Protocol~\ref{ptl:NLAND}. It allows Bob to abort during the protocol when he finds that Alice is cheating. We then propose Protocol~\ref{ptl:precompute1b} which includes checks from both sides to ensure that the average rate of cheating by any party is asymptotically vanishing (under the assumption of no physical noise).

When both parties are honest-but-curious, all the protocols are secure. An honest-but-curious party is one who follows the protocol while possibly making measurements which do not affect the final computation result. In our protocols, an honest-but-curious party does not learn anything about the other party's data, no matter whether the other party cheats or not.

The protocols with embedded checks in this paper allow aborts, circumventing the no-go theorem about two-party secure quantum evaluation of classical functions \cite{Lo97,bcs12}. See Sec.~\ref{sec2} below. In this paper we ignore the possible protocols that combine several one-time tables that potentially could have better security-efficiency tradeoff, and ignore the effects of physical noise.

The rest of the paper is organized as follows. Sec.~\ref{sec2} contains some introduction of the background. In Sec.~\ref{sec3} we introduce the quantum protocols for generating the one-time tables. Sec.~\ref{sec4} introduce two types of inequalities about guessing probabilities in some protocols for generating one-time tables, derived from a single inequality about semi-honest oblivious transfer protocols. Sec.~\ref{sec5} contains some discussions about physical implementations and about how to deal with physical noise. Sec.~\ref{sec6} contains the conclusion and some open problems.

\section{Preliminaries}\label{sec2}

On computing two-party classical functions with quantum circuits, Lo \cite{Lo97} studied the data privacy for publicly known classical functions with the output on one party only. Buhrman \emph{et al} \cite{bcs12} studied the security of two-party quantum computation for publicly known classical functions in the case that both parties know the outcome, although with some limitations in the security notions. These and other results in the literature \cite{Colbeck07} suggest that secure bipartite classical computing cannot be generally done by quantum protocols where the two parties have full quantum capabilities. In the current work, the protocols allow aborts in the quantum preprocessing (Bob may abort when he detects that Alice has cheated), and local randomness is used, so the scenario considered here does not fit into the assumptions in the works mentioned above.

Next, we introduce the simplest case in the one-time tables \cite{Beaver98}. It is also known as precomputed oblivious transfer, but note that our usage of the table is not for transferring a bit. It contains four bits: two distant bits $a$ and $b$, called ``input'' bits, and other two bits called ``output'' bits, which are $(a\cdot b)\oplus r$ and $r$ on the two parties, respectively, where $r$ is a uniformly random bit. (XOR is denoted as $\oplus$; AND is denoted as the $\cdot$ symbol.) Such correlation involving four bits is exactly that in the Popescu-Rohrlich type of nonlocal boxes \cite{Popescu1994,MAG06,PPK09}. Theoretically, the bipartite AND gate with distributed output on two distant input bits $a$ and $b$ can be computed while keeping both input bits completely private, with the help of a precomputed ideal one-time table of the nonlocal-AND type. Such one-time table has two locally-generated uniformly random bits $x$ and $y$ on Alice's and Bob's side, respectively, and also has $r'=(x\cdot y)\oplus r$ and $r$ on Alice's and Bob's side, respectively, where $r$ is a uniformly random bit. The steps for the bipartite AND-gate computation with distributed output are as follows:

1. Alice announces $a'=a \oplus x$. Bob announces $b'=b \oplus y$.

2. Each party calculates an output bit according to the one-time table and the received message. Alice's output is $(x\cdot b')\oplus r'=(x\cdot b')\oplus(x\cdot y)\oplus r$. Bob's output is $(a'\cdot b)\oplus r$.

The XOR of the two output bits is $(x\cdot b')\oplus(x\cdot y)\oplus r\oplus(a'\cdot b)\oplus r=a \cdot b$, while each output bit is a uniformly random bit when viewed alone, because $r$ is a uniformly random bit. Since the messages $a'$ and $b'$ do not contain any information about $a$ and $b$, the desired bipartite AND gate is implemented while $a$ and $b$ are still perfectly private.

With such capability above, it is easy to show that secure two-party classical computation can be performed \cite{Beaver98}. To see this, note that in the intermediate stages of the distributed classical computation, a logical data bit may be shared as the XOR of two bits on the two parties. The XOR gate between such logical data bits can be implemented by local XOR gates, while the AND gate with distributed output between such logical data bits can be implemented by local XOR gates and the nonlocal AND gates with distributed output discussed above.

Some notations are as follows. The random bits are unbiased and independent of other variables by default. We use the bit as the unit for information or entropic quantities.

\section{The quantum protocols for generating one-time tables}\label{sec3}

The Protocol~\ref{ptl:NLAND}, detailed in the table below, effectively computes an AND function on two remote classical bits from the two parties, with the output being a distributed bit, i.e. the XOR of two bits on the two parties. It is adapted from the semi-honest oblivious transfer protocol in \cite{CKS13,CGS16}, by changing the entangled state to a single-qutrit state, but using two states for each logical input value to recover the comparable level of security. The security of the inputs in Protocol~\ref{ptl:NLAND} is partial and comparable to that in the semi-honest oblivious transfer protocol in \cite{CKS13,CGS16}. Later we propose protocols that check the one-time tables generated from Protocol~\ref{ptl:NLAND}, to be used in the preprocessing stage in a bipartite classical or quantum computation task.

\begin{algorithm*}[htb]
\caption{A protocol containing two qutrits of communication for generating one-time tables with partial privacy}\label{ptl:NLAND}
\begin{flushleft}
\noindent{\bf Input:} A bit $x$ chosen by Alice before the protocol starts, and a bit $y$ chosen by Bob before the protocol starts. The distribution of both bits are uniformly random in the view of the other party or any outside party.\\
\noindent{\bf Output:} $r'=(x\cdot y)\oplus r$ on Alice's side, and $r$ on Bob's side, where $r$ is a random bit generated during the protocol, and it is unknown to any party before the protocol starts.\\
\noindent The input and output together form the one-time table.\\
\begin{enumerate}
\item Alice generates a uniformly random bit $t$. She prepares a qutrit in the state $\frac{1}{\sqrt{2}}(\ket{x}+(-1)^t \ket{2})$, where $x\in\{0,1\}$ is her input bit. She sends the prepared qutrit to Bob.\label{step1}
\item Bob receives qutrit from Alice. Bob generates a uniformly random bit $r$, which is to be regarded as his output bit. He performs the gate $(-1)^r \ketbra{0}{0} + (-1)^{y+r} \ketbra{1}{1}+ \ketbra{2}{2}$. He sends the qutrit to Alice.
\item Alice receives the qutrit from Bob. Alice measures the received qutrit in the basis $\{\frac{1}{\sqrt{2}}(\ket{x}+\ket{2}),\frac{1}{\sqrt{2}}(\ket{x}-\ket{2}),\ket{1-x}\}$. If the measurement outcome indicates that the state is the same as what she had sent to Bob in Step~\ref{step1}, she records her output bit as $0$, otherwise she records her output bit as $1$.
\end{enumerate}
\end{flushleft}
\end{algorithm*}

The correctness of the outputs of Protocol~\ref{ptl:NLAND} is easily verified. Alice's output bit, i.e. her measurement outcome in the last step is $r$ when $x=0$, or $y\oplus r$ when $x=1$, thus her output is equal to $(x\cdot y)\oplus r$. Bob's output bit is $r$.

In Protocol~\ref{ptl:NLAND}, Alice's input bit has partial privacy even for a cheating Bob, while Bob's input bit is secure for an honest-but-curious Alice, but is not secure at all for a cheating Alice.

The privacy of Alice's input bit $x$ can be quantified using the accessible information or the trace distance. The accessible information, i.e. the maximum classical mutual information corresponding to Bob's possible knowledge about Alice's input, is exactly $\frac{1}{2}$ bits, which happens to be equal to the Holevo bound in the current case. For a cheating Bob to get the maximum amount of information, his best measurement strategy in the current case is to measure in the computational basis $\{\ket{0},\ket{1},\ket{2}\}$. Alice's average density operator for input $x$ is $\frac{1}{2}(\ketbra{x}{x}+\ketbra{2}{2})$, where $x$ is $0$ or $1$. The trace distance of these two density operators is $\frac{1}{2}$, by direct calculation. (The trace distance is defined using $D(\rho,\sigma)=\frac{1}{2}\tr\vert\rho-\sigma\vert$, where $\vert A\vert\equiv\sqrt{A^\dag A}$.)
Thus, the probability that Bob guesses Alice's input bit correctly is $(1+\frac{1}{2})\cdot\frac{1}{2}=\frac{3}{4}$. This matches the probability of $\frac{3}{4}$ in \cite{CGS16} for Bob to guess correctly Alice's choice bit in a semi-honest oblivious transfer protocol. It can also be easily verified that
the maximum mutual information obtainable by Bob about Alice's choice bit in \cite{CGS16} is $\frac{1}{2}$, again by Bob measuring his qutrit in the computational basis. Note that with this particular computational-basis measurement, Bob cannot make the distributed output of the one-time table correct. In fact he has exactly $50\%$ chance to make it correct, the same chance as plain guessing. On the other extreme end, if Bob wants to make sure the distributed output of the one-time table is exactly correct, he cannot learn anything about Alice's input bit $x$, and the reason is in Prop.~\ref{prop:Holevo3} below. This implies that Alice could check for Bob's cheating by asking him to send her his input and output in some of the instances of Protocol~\ref{ptl:NLAND}, see Protocol~\ref{ptl:precompute1b} for details.

To learn about Bob's input bit, a cheating Alice may use the state $\frac{1}{\sqrt{2}}(\ket{0}+\ket{1})$ or the state $\frac{1}{\sqrt{2}}(\ket{0}-\ket{1})$. By measuring Bob's returned state in the basis $\{\frac{1}{\sqrt{2}}(\ket{0}+\ket{1}),\frac{1}{\sqrt{2}}(\ket{0}-\ket{1}),\ket{2}\}$, Alice may find out Bob's input bit $y$ with certainty. But in such case Alice has no effective input to speak of, and she does not know Bob's output bit $r$. Since $r$ is supposed to be randomly generated in the protocol, even if Alice chooses an input bit for herself later, she cannot determine her output bit for making the distributed output correct. Note that the average density operator for the two cheating states mentioned above is $\ketbra{0}{0}+\ketbra{1}{1}$, which is of the similar form as the density operators for Alice's inputs $0$ or $1$. Thus it can be easily calculated that if Bob wants to distinguish between the cases that whether Alice used the logical input value $x=0$ or used the cheating input state above, he would guess correctly with probability $\frac{3}{4}$, and the maximum mutual information obtainable by him about such distinction is $\frac{1}{2}$ bit. The same holds if $x=0$ is replaced with $x=1$.

The Protocol~\ref{ptl:NLAND} has two stages of communication. The total communication cost is two qutrits. In Sec.~\ref{sec5}, it will be mentioned that for photon-path encoding, the communication cost can be effectively reduced to two qubits, but this comes with the particular issue of how to make guarantee for single photons in optical encodings.

In the following we present protocols which check the one-time tables generated in Protocols~\ref{ptl:NLAND}. The Protocol~\ref{ptl:precompute1} has partial security for Alice and near-perfect security for Bob, while the Protocol~\ref{ptl:precompute1b} involves checking by both parties, and aims for near-perfect security for both parties.

\begin{algorithm*}[htb]
\caption{A partly-secure protocol for checking the one-time tables}\label{ptl:precompute1}
\begin{enumerate}
\item Alice and Bob perform many instances of Protocol~\ref{ptl:NLAND} (sequentially or in parallel) to generate some one-time tables, and exchange messages to agree on which instances were successfully implemented experimentally. Suppose $M$ one-time tables were implemented. The one-time tables labeled by $j$ has inputs $a_j$ and $b_j$, and outputs $e_j$ and $f_j$.
\item Bob randomly selects $K$ integers in $\{1,\cdots,M\}$, which are labels for which one-time table. He tells his choices to Alice. The integer $K$ satisfies that $M-K$ is an upper bound on the number of required one-time tables in the main bipartite computing task, and the ratio $\frac{K}{M}$ is related to the targeted security level of the overall computation.
\item Alice sends the bits $a_j$ and $e_j$ to Bob for all chosen labels $j$.
\item For any chosen label $j$, Bob checks whether $a_j$ and $e_j$ satisfy that $a_j \cdot b_j=e_j \oplus f_j$. If the total number of failures is larger than some preset number of Bob's (e.g. a small constant, or a small constant times $K$), he aborts the protocol, or restarts the protocol to do testing on a new batch of instances of Protocol~\ref{ptl:NLAND} if the two parties still want to perform some secure two-party computation. Otherwise, the remaining one-time tables are regarded as having passed the checking and will be used later in the two-party computing task. They may repeat the steps above to prepare more one-time tables on demand.
\end{enumerate}
\end{algorithm*}

In Protocol~\ref{ptl:precompute1}, Alice's input bit has partial privacy, which is the same as in the analysis of Protocol~\ref{ptl:NLAND} above. When the ratio $\frac{K}{M}$ is near one, the nonlocal correlations in the remaining unchecked one-time tables can be regarded as almost surely correct. This is because of Bob's checking. We require Alice to be weakly cooperating, that is, she does not cheat in some of the batches of instances, since otherwise no one-time table may pass the test. Some degree of weak cooperation is required for two parties to perform a computation anyway, and the above assumption of Alice has no effect on the data security of any party when Bob satisfies the assumption below, thus we may ignore the assumption above and just state the following assumption on Bob as the requirement of our protocols. In the following we assume that Bob is \emph{conservative}, which means that he values the privacy of his data higher than the possibility to learn Alice's data. Operationally this implies Bob would do the checking as specified in our protocols. For an honest-but-curious Alice, the resulting correlation is correct, and she does not learn anything about Bob's input bit $y$ (using the notations in Protocol~\ref{ptl:NLAND}, same below). In the following we discuss the case that Alice cheats.

If Alice cheats and gets at least partial information about Bob's input bit $y$, the state sent from Alice to Bob must be different from what is specified in the protocol; some of her best choices of the states for cheating are mentioned previously. To pass Bob's test while learning about Bob's input $y$, she should know both $y$ and $r$, or know both $y$ and $y \oplus r$. (The two conditions are equivalent in the exact case, but not necessarily equivalent in the partial-information case.) In the following, let $I^{\cal M}_y$ denote the classical mutual information learnable by Alice about Bob's bit $y$ if she uses the measurement $\cal M$ on the received state (possibly a POVM measurement), in an instance of Protocol~\ref{ptl:NLAND}. The $I^{\cal M}_r$ and $I^{\cal M}_{y\oplus r}$ are defined similarly. In our applications in the protocols in this paper, we always assume that the prior distribution of $y$ and $r$ are the uniform distribution as long as we say that ``Bob is honest''. But by looking into the proofs of Lemma~\ref{lemma1} and Prop.~\ref{prop1} below, such requirement is not actually necessary.

Before stating the Prop.~\ref{prop1} which is directly relevant to Bob's security in the protocols in this paper, we first state a technical lemma, proved in Appendix~\ref{app:prooflemma1}.

\begin{lemma}\label{lemma1}
Let $X=\{x_i,p_i\}_{i=1}^4$ be a classical random variable containing four possible messages $x_i$, each with some probability $p_i$. Let two bits $r$ and $y$ be the labels representing the message. Suppose $X$ is encoded using one of four pure quantum states of a qutrit as follows
\bea
r=0,\, y=0:\quad && a\ket{0}+b\ket{1}+c\ket{2},\notag\\
r=0,\, y=1:\quad && a\ket{0}-b\ket{1}+c\ket{2},\notag\\
r=1,\, y=0:\quad && -a\ket{0}-b\ket{1}+c\ket{2},\notag\\
r=1,\, y=1:\quad && -a\ket{0}+b\ket{1}+c\ket{2},\label{eq:newstate0}
\eea
where $a,b,c\in\mathbb{R}$. Then the maximum amount of classical mutual information about $X$ that can be obtained by a party performing a POVM measurement on the qutrit is not greater than $1$ bit.
\end{lemma}

The proof of the following Prop.~\ref{prop1} is in Appendix~\ref{app:proofp1}. The reason why we implicitly use a density operator $\sigma_A$ in the assumption instead of using Alice's pure initial state on a joint system is given in the proof.

\bpp\label{prop1}
In Protocol~\ref{ptl:NLAND} where Bob is honest but Alice may cheat, the following inequalities hold:
\bea
I^{\cal M}_y+I^{\cal M}_r &\le& 1,\label{eq:info1}\\
I^{\cal M}_y+I^{\cal M}_{y\oplus r} &\le& 1,\label{eq:info2}\\
I^{\cal M}_y+\max(I^{\cal M}_r,I^{\cal M}_{y\oplus r}) &\le& 1.\label{eq:info3}
\eea
where the two $\cal M$ are the same in each equation. All the quantities on the left-hand-sides are also dependent on Bob's received state $\sigma_A$. It is effectively prepared by Alice, and is a mixed state on a qutrit, and the two $\sigma_A$ are the same in each equation. (We abbreviate the symbol $\sigma_A$.)
\epp

In the following we introduce Prop.~\ref{prop:probability}, which is not explicitly used later in this paper, but since it does not require the same measurement for learning about $y$ or $r$, it is quite different from Prop.~\ref{prop1} and its extreme case (one probability being $1$ and the other being $\frac{1}{2}$) is helpful for understanding Theorem~\ref{thm1} below. It also presents a small improvement over the corresponding result in \cite{CGS16}. In other words, there should be a corresponding inequality for semi-honest quantum oblivious transfers which is slightly tighter than the form in \cite{CGS16}.

\bpp\label{prop:probability}
In Protocol~\ref{ptl:NLAND} where Bob is honest but Alice may cheat, for a fixed (possibly cheating) input state of Alice, let the probability that Alice guesses Bob's bit $r$ correctly as $P_r$, and the probability that she guesses Bob's bit $y$ correctly as $P_y$, then
\bea
(P_r-\frac{1}{2})^2+(P_y-\frac{1}{2})^2\le \frac{1}{4}.\label{eq:guessingprob}\\
(P_{y\oplus r}-\frac{1}{2})^2+(P_y-\frac{1}{2})^2\le \frac{1}{4}.\label{eq:guessingprob2}
\eea
\epp

The proof of Prop.~\ref{prop:probability} is in Appendix~\ref{app:proofp2}. In Appendix~\ref{app:example}, we provide some examples, some of which satisfy the equality in some inequalities Eqs.~\eqref{eq:info1} and \eqref{eq:guessingprob}.

The probability that Alice passes Bob's test at a particular instance is related to the $\max(I^{\cal M}_r,I^{\cal M}_{y\oplus r})$ in Eq.~\eqref{eq:info3}. When the probability of passing approaches $1$, such maximum approaches $1$, then it must be that one of them approaches $1$. Then, Prop.~\ref{prop1} implies that Alice can learn almost nothing about $y$ if she measured in the same basis, but in fact a cheating Alice knows which instances are remaining and will not be checked later, so she can choose to do any measurement on the received states in these remaining instances. Such measurement may not be the same as $\cal M$ in the other term in Eq.~\eqref{eq:info3}. This implies that Eq.~\eqref{eq:info3} alone is not sufficient for proving the security of Protocol~\ref{ptl:precompute1}. We note that Prop.~\ref{prop:probability} does not require the same measurements for learning about $y$ or $r$, and the extreme case in the result of Prop.~\ref{prop:probability} explains the security in the corresponding case of Protocol~\ref{ptl:precompute1}, but for the intermediate cases we still need to obtain some quantitative relation in terms of information quantities rather than probabilities. Although it is possible to study the mutual information tradeoffs for different measurements in a single copy of Protocol~\ref{ptl:NLAND}, the joint measurements across copies present challenges for further study. This is why in the following we study the Holevo bounds instead.

In the following we consider the Holevo bounds for the classical mutual information about $y$ or $r$, or $y\oplus r$. Under the condition that $y$ is uniformly distributed on the two-element set $\{0,1\}$, the Holevo bound for information about $y$ is
\bea\label{eq:defHolevo}
\chi_y=S(\rho)-\frac{1}{2}\sum_{j=0}^1 S(\rho_j),
\eea
where $\rho_j$ is the density operator that Alice receives from Bob for the case of $y=j$ after Pauli corrections determined by Bob's sent bit, and $\rho=\frac{1}{2}(\rho_0+\rho_1)$. The $S$ represents the von Neumann entropy.
The density operators for $y=0$ and $y=1$ are given in Eqs.~\eqref{eq:y0} and \eqref{eq:y1}, and it follows that the density operator averaged over $y$ is $\rho=a^2\ketbra{0}{0}+b^2\ketbra{1}{1}+c^2\ketbra{2}{2}$. Therefore, noting that $a^2+b^2+c^2=1$, we have
\bea
\chi_y=-a^2\log_2 a^2-b^2\log_2 b^2-c^2\log_2 c^2\notag\\
+c^2\log_2 c^2+(1-c^2)\log_2(1-c^2)\notag\\
=-a^2\log_2 a^2-b^2\log_2 b^2+(1-c^2)\log_2(1-c^2).\label{eq:chiy}
\eea
We can similarly define $\chi_r$ and $\chi_{y\oplus r}$. By a similar argument, we obtain
\bea
\chi_r=-a^2\log_2 a^2-c^2\log_2 c^2+(1-b^2)\log_2(1-b^2),\notag\\
\label{eq:chir}\\
\chi_{y\oplus r}=-b^2\log_2 b^2-c^2\log_2 c^2+(1-a^2)\log_2(1-a^2).\notag\\
\label{eq:chiyxorr}
\eea
This gives rise to the following result.

\bpp\label{prop:Holevotradeoff}
The following statements hold for Protocol~\ref{ptl:NLAND} where Bob is honest but Alice may cheat.\\
(i) Suppose $\delta=1-\chi_r$ is in the range $[0,0.5)$, the following relations about Holevo quantities hold:
\bea
\chi_y\le h(\delta),\label{eq:chitradeoff1}\\
\chi_{y\oplus r}\le h(\delta),\label{eq:chitradeoff2}
\eea
where $h(\delta)\equiv -(1-\delta)\log_2 (1-\delta)-\delta\log_2 \delta$.\\
(ii) Suppose $\delta'=1-\chi_{y\oplus r}$ is in the range $[0,0.5)$, the following relations about Holevo quantities hold:
\bea
\chi_r\le h(\delta'),\label{eq:chitradeoff3}\\
\chi_y\le h(\delta').\label{eq:chitradeoff4}
\eea
\epp

The proof of Prop.~\ref{prop:Holevotradeoff} is in Appendix~\ref{app:proofp3}. Now we are in a position to obtain some assertion about the security of Protocol~\ref{ptl:precompute1}.

\bt\label{thm1}
In Protocol~\ref{ptl:precompute1}, honest Bob's input is asymptotically secure.
\et

The proof of Theorem~\ref{thm1} is in Appendix~\ref{app:proofthm1}. The proof contains an estimate of the cost overhead ratio due to checks, under some reasonable assumption about how to predict future failure rates from tested instances of Protocol~\ref{ptl:NLAND}. The overhead ratio can be small compared with the number of one-time tables to be prepared. In Appendix~\ref{app:num} we present some numerical results about the Holevo quantities and mutual information arising from Protocol~\ref{ptl:NLAND}.

To improve Alice's security in the protocol above, we propose the following Protocol~\ref{ptl:precompute1b}, in which Alice also does some checking about Bob's behavior.

\begin{algorithm*}[htb]
\caption{A protocol for checking the one-time tables by both parties}\label{ptl:precompute1b}
\begin{enumerate}
\item Alice and Bob perform many instances of Protocol~\ref{ptl:NLAND} to generate some one-time tables, and exchange messages to agree on which instances were successfully implemented experimentally. Suppose $M$ one-time tables were implemented. The one-time tables labeled by $j$ has inputs $a_j$ and $b_j$, and outputs $e_j$ and $f_j$.
\item (The steps 2 to 4 can be done concurrently with the steps 5 to 7.) Bob randomly selects $K_B$ integers in $\{1,\cdots,M\}$, which are labels for which one-time table. He tells his choices to Alice.
\item Alice sends the bits $a_j$ and $e_j$ to Bob for all chosen labels $j$.
\item For any chosen label $j$, Bob checks whether $a_j$ and $e_j$ satisfy that $a_j \cdot b_j=e_j \oplus f_j$. If the total number of failures is larger than some preset number of Bob's (e.g. $0$, or a small constant times $M$), he aborts the protocol, or asks Alice to restart the protocol to do testing on a new batch of instances of Protocol~\ref{ptl:NLAND} if the two parties still want to perform some secure two-party computation.
\item Alice randomly chooses $K_A$ integers in $\{1,\cdots,M\}$, and tells Bob her choices. The chosen set of integers may overlap with the set chosen by Bob.
\item Bob sends the bits $b_j$ and $f_j$ to Alice for the chosen labels $j$.
\item For any chosen label $j$, Alice checks whether $a_j \cdot b_j=e_j \oplus f_j$ holds. If the total number of failures is larger than some preset number of Alice's, she aborts the protocol, or asks Bob to restart the protocol if needed.
\item The remaining one-time tables are regarded as having passed the checking and will be used later in the two-party computing task. They may repeat the steps above to prepare more one-time tables on demand.
\end{enumerate}
\end{algorithm*}

On the security of honest Alice's input bits in Protocol~\ref{ptl:precompute1b} when Bob may possibly cheat, there is an analogue of Theorem~\ref{thm1} for Alice instead of Bob, see Theorem~\ref{thm2} below. To draw an analogy to the analysis of Protocol~\ref{ptl:precompute1}, note that the output bits of Protocol~\ref{ptl:NLAND} can alternatively be written as $r'$ on Alice's side and $(x\cdot y)\oplus r'$ on Bob's side, respectively, where $r'$ is a uniformly random bit, and is related to the $x,y,r$ by the equation $r'=(x\cdot y)\oplus r$.

We model Bob's operations including possible measurement in Protocol~\ref{ptl:NLAND} using a unitary $\cU: \cH_A \ox \cH_B \ox \cH_E \rightarrow \cH_{A'} \ox \cH_B \ox \cH_F$, followed by some measurement on (some subsystem of) $\cH_B\ox \cH_F$. The $A'$ is Alice's output system, and the $B$ and $F$ combined is Bob's output system, denoted as $B'$ below. Alice's initial state is in system $A$, and Bob's initial values of $y$ and $r$ are encoded into computational-basis quantum states in the input system $B$. The $B$ does not contain other subsystems, and all other ancilla is in system $E$. Note that the actual operations may involve measurements before other unitary gates, but we always defer the measurements to get equivalent outcomes; Alice's system may be partially measured by Bob before being sent to Alice, and in such case we model Bob's such measurement as a unitary on a larger system by including the measurement apparatus and any systems recording the results, as well as any possible ancillary systems into $\cH_E$, so as to make $\cU$ unitary. We use $M(\cU)$ to refer to the measurements in the inequalities below, where the first part of the measurement is the unitary $\cU$, and the latter part of the overall measurement are all local unitaries on $\cH_{A'}$ and $\cH_{B'}$ followed by local (POVM) measurements in the local subsystems mentioned below.

Let $I^{M(\cU)}(x,B')$ be the accessible information (maximal classical mutual information) learnable by Bob about Alice's input $x$ using the unitary $\cU$ followed by an arbitrary measurement on Bob's output system $B'$. The ``maximal'' above is maximizing over the measurement on $B'$ after the fixed unitary $\cU$. The other type of information to be considered is the amount of classical information learnable by Alice about the value of $q\equiv t \oplus (x\cdot y)\oplus r$ given the value of $x$, where $t$ is Alice's random bit generated locally in Protocol~\ref{ptl:NLAND}. The condition ``given the value of $x$'' appears because Alice's final measurement basis depends on the value of $x$ but is otherwise fixed. Such basis is known to her but is not explicit in the state sent to Bob. Since Bob may choose the value of $y$ when later asked to send Alice the value of $y$ and $r$ to be checked, we have to separately consider two quantities: $I^{M(\cU)}_{t \oplus r \vert x}$ which is the amount of classical mutual information about $t\oplus r$ given $x$, and $I^{M(\cU)}_{t \oplus x\oplus r \vert x}$ which is the amount of classical mutual information about $t\oplus x\oplus r$ given $x$. They correspond to the cases $y=0$ and $y=1$, respectively. Note that since we consider the case that Bob may possibly cheat, the $y$ and $r$ here are understood as Bob's initial variables before his unitary $\cU$, but should not be understood as that Bob does exactly the operations corresponding to $y$ and $r$ according to the description of Protocol~\ref{ptl:NLAND}.

The Proposition~\ref{prop:Holevo3} below is for proving Theorem~\ref{thm2}, which is about honest Alice's security in Protocol~\ref{ptl:precompute1b}. Note that for each value of $x$, there is a measurement on $A'$ for learning about the value of $q\equiv t \oplus r \oplus (x\cdot y)$ given $x$. The Prop.~\ref{prop:Holevo3} uses the Holevo bound for each of the two measurements on $A'$. Note that for the unchecked instances of Protocol~\ref{ptl:NLAND}, Bob may use a different measurement on his output system $B'$ than what he uses on his output system in the checked instances, but since we do allow arbitrary local measurements in the local subsystems in the inequalities in Prop.~\ref{prop:Holevo3} below, such issue is actually taken into consideration.

\bpp\label{prop:Holevo3}
Suppose $\delta\in(0,0.1)$ is a small constant. The statement holds for Protocol~\ref{ptl:NLAND} where Alice is honest in the initial stage (up to sending of the prepared state to Bob) but Bob may cheat. If
\bea\label{eq:chisum3a}
\frac{1}{2}\left[I^{M(\cU)}_{t \oplus r \vert{x=0},A'} + I^{M(\cU)}_{t \oplus r \vert{x=1},A'}\right]>1-\delta,
\eea
then
\bea\label{eq:tradeoff7}
I^{M(\cU)}_{x,B'}=O(\delta^{1/4}\log\frac{1}{\delta})
\eea
for sufficiently small $\delta$ (but smaller than $0.1$ anyway), where there is no limit to the dimension of the ancilla space ${\cal H}_E$ used by $\cU$ as long as it is finite.
\epp

The proof of Prop.~\ref{prop:Holevo3} is in Appendix~\ref{app:proofp4}. The following Theorem~\ref{thm2} concerns honest Alice's security in Protocol~\ref{ptl:precompute1b}, while honest Bob's security is guaranteed using the same arguments as in the proof of Theorem~\ref{thm1}.

\bt\label{thm2}
In Protocol~\ref{ptl:precompute1b}, honest Alice's input is asymptotically secure.
\et

The proof of Theorem~\ref{thm2} is in Appendix~\ref{app:proofthm2}. It also contains an estimate of the cost overhead ratio, which is similar to that in the proof of Theorem~\ref{thm1}. In Protocol~\ref{ptl:precompute1b}, if any one party is conservative, his (her) data privacy is guaranteed. Partly due to the possible aborts, it actually suffices to assume either one of the parties is conservative in Protocol~\ref{ptl:precompute1b}, since then the other party might as well be conservative to reach a better security level for himself (herself).

\section{Inequalities about guessing probabilities in some protocols for generating one-time tables}\label{sec4}

In this section, we first introduce an inequality from \cite{CGS16} about guessing probabilities of the two parties in generic protocols without aborts for semi-honest quantum oblivious transfer. In oblivious transfer, Bob transfers one of two bits to Alice, and is oblivious as to which bit he transferred. Semi-honest oblivious transfer are those oblivious transfers in which Alice knows one of Bob's bits with certainty. We use the inequality to derive two types of inequalities about guessing probabilities, each for a class of protocols for generating one-time tables. The derivation for the second type is somewhat unexpected. We adopt the following notation: in an oblivious transfer protocol, suppose Alice's choice bit is $a$, and Bob's bits to be transferred are $x_0$ and $x_1$.

\bpp\label{prop3}
(\cite[Theorem 1]{CGS16}) Let $P^\star_{\rm Bob}$ denote the probability that Bob can guess honest-Alice's choice bit $a$ correctly. Let $P^\star_{\rm Alice}$ be the maximum probability over $a\in\{0,1\}$ that cheating-Alice can guess $x_{\bar a}$ correctly while knowing $x_a$ with certainty. ($\bar a=1-a$.) Then for any oblivious transfer protocol without aborts satisfying the above (implying that the protocol is for semi-honest oblivious transfer), the following inequality holds:
\begin{equation}\label{ineqP1}
2P^\star_{\rm Bob}+P^\star_{\rm Alice}\ge 2.
\end{equation}
\epp

In the following we adopt the same notations as in Protocol~\ref{ptl:NLAND}: Alice's and Bob's input bits are $x$ and $y$, respectively; Alice's output bit is $r'=(x\cdot y)\oplus r$, and Bob's output bit is $r$. A ``correct protocol'' refers to that Alice can obtain the desired output for $x=0$ and for $x=1$ by choosing suitable inputs and operations according to $x$.

\bt\label{thm3}
(i) In any correct protocol without aborts for generating one-time tables, let $P_A$ be the maximum probability over $a\in\{0,1\}$ that cheating-Alice guesses correctly her output for $x=a$ while learning her output with certainty in the case $x=1-a$, and let $P_B$ be the probability that Bob guesses correctly honest-Alice's input bit $a$. Then the following inequality holds:
\begin{equation}\label{ineqP2}
2P_B+P_A\ge 2.
\end{equation}

(ii) Consider those protocols for generating one-time tables in which cheating-Alice can learn Bob's input $y$ with certainty, and there are no aborts. Let $P_{Ar}$ denote the probability that cheating-Alice guesses correctly the $r$ when her operations (all quantum and classical operations including possible state preparation and measurements, same below) are such that she learns $y$ with certainty; let $P_{Ay}$ denote the probability that cheating-Alice guesses correctly the $y$ when her operations are such that she learns $r$ with certainty. Let $P'_B$ be the probability that Bob guesses correctly whether Alice's operations are for learning $y$ or learning $r$. Then the following inequality holds:
\begin{equation}\label{ineqP3}
2P'_B+\max\{P_{Ar},P_{Ay}\}\ge 2.
\end{equation}
The similar inequality holds when $r$ is replaced with $y\oplus r$.
\et
\bpf
(i) The protocol for generating one-time tables can be viewed as a protocol for oblivious transfer of the bits $r$ and $y\oplus r$. These bits are Alice's output bits in the protocol: the $r$ is for input $x=0$, and the $y\oplus r$ is for input $x=1$. Thus we may directly apply Prop.~\ref{prop3} and obtain the inequality~\eqref{ineqP2}.

(ii) Under the assumption that cheating-Alice can learn Bob's input $y$ with certainty, the protocol for generating one-time tables can be viewed as a protocol for oblivious transfer of the bits $y$, and $r$ in the case $x=0$ (or $y\oplus r$ in the case $x=1$). Thus we may apply Prop.~\ref{prop3} and obtain the inequality~\eqref{ineqP3}.
\epf

Note that the last ``Alice's operations'' in the statement of Theorem~\ref{thm3}~(ii) may often refer to Alice's initial state preparations, which is the case in Protocol~\ref{ptl:NLAND}, where what Alice wants to do (to cheat or using an honest input, e.g. $0$) is entirely determined by her prepared initial state and independent of her last measurement.

An extreme case for the equality to be reached in Eq.~\eqref{ineqP2} is achieved by Protocol~\ref{ptl:NLAND}, which is $P_A=\frac{1}{2}$ (which means no information), $P_B=\frac{3}{4}$, see the discussion about Protocol~\ref{ptl:NLAND} in Sec.~\ref{sec3}. An extreme case for the equality to be reached in Eq.~\eqref{ineqP3} is achieved by Protocol~\ref{ptl:NLAND}, which is $P_{Ar}=P_{Ay}=\frac{1}{2}$, $P'_B=\frac{3}{4}$.

\section{Discussions}\label{sec5}

{\bf 1. Comparison with a previously proposed protocol.}

Preliminary studies show that the security characteristics of Protocol~\ref{ptl:NLAND} is similar to that of Protocol 1 in \cite{Yu19}, the latter involving somewhat higher communication cost, i.e. sending two qubits in both directions. In trying to compare the protocols, we have discovered a slightly improved cheating strategy for Alice in Protocol 1 in \cite{Yu19}, and the comparison just mentioned is made after such changes. But we suspect that the feasibility of generalization to qudits may be different for the two protocols. We leave the details to further study.

{\bf 2. On physical implementations of Protocol~\ref{ptl:NLAND}.}

The Protocol~\ref{ptl:NLAND} involves sending of qutrits. Because our protocols do not require the two parties to be at remote positions, the qutrit in the protocol could be implemented by solid-state physical systems. The two parties take turns to operate on the physical systems. On the other hand, let us consider optical encoding when the two parties are allowed to be distant from each other. Since the polarization space of a photon is only two-dimensional, the path encoding could be a possible candidate. In using the path degree of freedom, note that Bob only needs the subspace spanned by $\{\ket{0},\ket{1}\}$, hence the effective communication cost is only two qubits, but a drawback is that under this and many other optical encodings, Bob needs to check whether Alice had used single photons. Other potential optical degrees of freedom include time-of-arrival, or orbital angular momentum. Combinations of them (including polarization) could also be considered.

{\bf 3. Dealing with noise and errors.}

While Bob's gates in a previously proposed protocol involving sending two qubits \cite{Yu19} are Clifford gates,
Bob's gates in the current Protocol~\ref{ptl:NLAND} are not Clifford operators. This means that we can not straightforwardly apply fault-tolerant computation techniques here, but the gates here are very simple, so there would likely be some encoding that allow effective fault-tolerant implementation of the gates. Note that this might not be equivalent to the fault-tolerance of the entire Protocol~\ref{ptl:NLAND}. There is also the problem of extending fault-tolerance to the entire check-based generation of one-time tables, or even to the entire two-party classical or quantum computation. We leave such problems to later study.

\section{Conclusion}\label{sec6}

We have proposed a qutrit-based quantum protocol for generating a certain type of classical correlations (a special case of the one-time tables \cite{Beaver98}, the same correlation as in the Popescu-Rohrlich nonlocal box) with partial privacy, and proposed protocols for checking the generated correlations, and one of the protocols achieves check-based asymptotic information-theoretic security for both parties in the generated one-time tables. An estimate of the cost overhead ratio due to checks is also presented in the proof of some theorems, under some reasonable assumption about how to predict future failure rates from tested instances of a subprocedure. Our methods are not direct implementation of nonlocal boxes, since the standard notion of nonlocal boxes involves some instantaneous effect, while our methods require some time and communication cost. As a side result, we have found an inequality about guessing probabilities, which improves upon a corresponding result in \cite{CGS16}. We have also obtained two other types of inequalities about guessing probabilities in some general classes of quantum protocols without aborts for generating one-time tables, from a single inequality about guessing probabilities in semi-honest quantum oblivious transfer. The methods of using the one-time tables in bipartite secure (interactive) classical or quantum computation tasks are known in the literature (e.g.~\cite{Beaver98}), but we think the issues with using imperfect one-time tables have not been thoroughly studied. We leave the applications or extensions of our protocols to future study. On improving or using the current set of protocols using qutrits, some open problems include: how to achieve fault-tolerance; design of experimental schemes; extensions of the protocols for implementing other nonlocal correlations.

\smallskip
\section*{Acknowledgments}

This research is supported by the National Natural Science Foundation of China (No. 11974096, No. 61972124, No. 11774076, and No. U21A20436), and the NKRDP of China (No. 2016YFA0301802).

\linespread{1.0}
\bibliographystyle{unsrt}
\bibliography{homo}

\begin{appendix}

\section{Proof of Lemma~\ref{lemma1}}\label{app:prooflemma1}

\bpf
Let $\cal M$ denote the POVM measurement on the qutrit, and let $I^{\cal M}_{y,r}$ denote the classical mutual information between the distribution of measurement outcomes of $\cal M$ and the distribution of the input, described using the bits $y$ and $r$. We shall use the Holevo bound to prove that $I^{\cal M}_{y,r}\le 1$. For a given set of encoding density operators $\rho_j$ and associated probabilities $p_j$, the Holevo bound is an upper bound for the accessible information, the latter being the largest classical mutual information under all possible measurements. It is also called the Holevo $\chi$ quantity. It is defined as
\bea
\chi=S(\rho)-\sum_j p_j S(\rho_j),\label{eq:Holevo_general}
\eea
where $\rho=\sum_j p_j \rho_j$, and $S\equiv -\tr\rho\log_2\rho$ is the von Neumann entropy. Our proof approach is to map the four pure states in Eq.~\eqref{eq:newstate0}, which are in a $3$-dimensional Hilbert space, to possibly mixed states in a $2$-dimensional Hilbert space. The entropy of the average state in the $2$-dimensional Hilbert space is not greater than $1$ bit. Thus the Holevo $\chi$ quantity is at most $1$ bit, proving that the accessible information for measuring in the $2$-dimensional Hilbert space is at most $1$ bit. But what we wanted to prove is that the accessible information for measuring in the original $3$-dimensional Hilbert space is at most $1$ bit. Thus we want to show that the measurement statistics are indeed the same in the two spaces, for measuring the four states or their probabilistic mixtures.

The explicit mapping we have found turns out to satisfy that the four original states are mapped to fixed pure states, while the POVM measurement is changed according to the original state, so that the measurement statistics are the same. The density operators for the four fixed target pure states are
\bea
r=0,\, y=0:\quad && \frac{1}{2}[\mathbb{I}_2+\frac{1}{\sqrt{3}}(\sigma_x+\sigma_y+\sigma_z)],\notag\\
r=0,\, y=1:\quad && \frac{1}{2}[\mathbb{I}_2+\frac{1}{\sqrt{3}}(-\sigma_x-\sigma_y+\sigma_z)],\notag\\
r=1,\, y=0:\quad && \frac{1}{2}[\mathbb{I}_2+\frac{1}{\sqrt{3}}(\sigma_x-\sigma_y-\sigma_z)],\notag\\
r=1,\, y=1:\quad && \frac{1}{2}[\mathbb{I}_2+\frac{1}{\sqrt{3}}(-\sigma_x+\sigma_y-\sigma_z)],\label{eq:newstate2}
\eea
where $\sigma_x,\sigma_y,\sigma_z$ are the qubit Pauli operators. Each density operator in Eq.~\eqref{eq:newstate2} is of the form $\frac{1}{2}(\mathbb{I}_2+\vec a \cdot \vec \sigma)$ with $\vert \vec a\vert=1$, hence it represents a pure state. The four points corresponding to these states actually form a regular tetrahedron on the Bloch sphere. Recall that for a density operator $\rho$, and a POVM element $A$ (satisfying that $A\ge 0$), the probability that a measurement outcome corresponding to POVM element $A$ appears is $\tr(A\rho)$.
For the receiver Alice to learn more information, she should use rank-$1$ POVM elements, since if there is a POVM element with rank greater than $1$, she could split it into some POVM elements of rank $1$, and the information she learns does not decrease. Hence, in the following we assume all POVM elements have rank $1$. For each density operator $\rho_j$ in \eqref{eq:newstate2}, we claim that there exist positive semi-definite operators $E_x, E_y, E_z$ [see Eq.\eqref{eq:povmE} below] such that $\tr(E_x\rho_j)$, $\tr(E_y\rho_j)$ and $\tr(E_z\rho_j)$ are equal to $\tr(\sigma'_x\tau_j)$, $\tr(\sigma'_y\tau_j)$ and $\tr(\sigma'_z\tau_j)$, respectively, where $\tau_j$ refers to the density operator in the $3$-dimensional Hilbert space corresponding to a pure state in \eqref{eq:newstate0}, and $\sigma'_x, \sigma'_y, \sigma'_z$ are operators in the $3$-dimensional Hilbert space listed as follows:
\bea
\sigma'_x&=&\ketbra{0}{1}+\ketbra{1}{0},\notag\\
\sigma'_y&=&\ketbra{0}{2}+\ketbra{2}{0},\notag\\
\sigma'_z&=&\ketbra{1}{2}+\ketbra{2}{1}.
\eea
Since POVM elements are Hermitian nonnegative operators, we may assume that every POVM element in the $3$-dimensional Hilbert space is a linear combination of the form
\bea
M_j&=& u_j\sigma'_x+v_j\sigma'_y+w_j\sigma'_z+i \a_j (\ketbra{0}{1}-\ketbra{1}{0})\notag\\
&&+i \b_j (\ketbra{0}{2}-\ketbra{2}{0})+i \g_j (\ketbra{1}{2}-\ketbra{2}{1})\notag\\
&&+f_j\ketbra{0}{0}+g_j\ketbra{1}{1}+h_j\ketbra{2}{2},\label{eq:linearcomb0}
\eea
where $j$ is the label for which POVM element, and the coefficients $\a_j,\b_j,\g_j,u_j,v_j,w_j\in\mathbb{R}$, $f_j,g_j,h_j\ge 0$. We also have $M_j\ge 0$. Since the states in \eqref{eq:newstate0} are real, it can be verified that the three terms with imaginary coefficients in \eqref{eq:linearcomb0} contribute zero to the probability $\tr(M_j \ketbra{\psi}{\psi})$ where $\ket{\psi}$ is a state in \eqref{eq:newstate0}. The completeness relation satisfied by $\{M_j\}$ is $\sum_j M_j=\mathbb{I}_3$. Now, we define
\bea
M'_j&=& u_j\sigma'_x+v_j\sigma'_y+w_j\sigma'_z+\notag\\
&&f_j\ketbra{0}{0}+g_j\ketbra{1}{1}+h_j\ketbra{2}{2},\label{eq:linearcomb}
\eea
where $u_j,v_j,w_j,f_j,g_j,h_j$ are the same as above, so we have $\sum_j M'_j=\mathbb{I}_3$, by considering the real part in the original completeness condition. From the condition $M_j\ge 0$, we obtain $M'_j\ge 0$, from the following argument: define the operator $M''_j$ to be the same as $M_j$ except for that the imaginary terms (which are off-diagonal) are multiplied by the factor $(-1)$. Then the condition $M_j\ge 0$ is equivalent to $M''_j\ge 0$, since $\bra{\psi}M_j\ket{\psi}=\bra{\tilde\psi}M_j\ket{\tilde\psi}$ where $\ket{\tilde\psi}$ is the complex conjugate of $\ket{\psi}$. Hence $M'_j=\frac{1}{2}(M_j+M''_j)\ge 0$. The text below Eq.~\eqref{eq:linearcomb0} implies that $\tr(M'_j \ketbra{\psi}{\psi}) = \tr(M_j \ketbra{\psi}{\psi}),\,\forall j$, for $\ket{\psi}$ being a state in \eqref{eq:newstate0}. Hence the measurement statistics from the POVM $\{M_j\}$ is completely the same as that from the POVM $\{M'_j\}$, and we use the latter set of POVM elements in the derivation below. We may find a POVM element in the $2$-dimensional Hilbert space corresponding to $M'_j$ as follows:
\bea
E_j=(a^2 f_j + b^2 g_j + c^2 h_j)\mathbb{I}_2+\notag\\
\sqrt{3}(ab u_j\sigma_x+ ac v_j \sigma_y+ bc w_j\sigma_z),\label{eq:povmE}
\eea
Since $u_j\le \sqrt{f_j g_j}$, which is from $M'_j\ge 0$, we have $\sqrt{3}ab u_j\le \sqrt{3a^2 f_j b^2 g_j}$, hence $(\sqrt{3}ab u_j)^2\le 3a^2 f_j b^2 g_j \le 2a^2 f_j b^2 g_j+[(a^2 f_j)^2 + (b^2 g_j)^2]/2$. Thus, we obtain
\bea
&&(\sqrt{3}ab u_j)^2+(\sqrt{3}ac v_j)^2+(\sqrt{3}bc w_j)^2\notag\\
&&\le (a^2 f_j + b^2 g_j + c^2 h_j)^2.
\eea
Hence $E_j\propto \mathbb{I}_2+\vec a \cdot \vec \sigma$ with $\vert \vec a\vert\le 1$, thus $E_j\ge 0$. The completeness relation satisfied by $\{M'_j\}$ is $\sum_j M'_j=\mathbb{I}_3$, and this implies $\sum_j f_j=\sum_j g_j=\sum_j h_j=1$, and $\sum_j u_j=\sum_j v_j=\sum_j w_j=0$. Together from the normalization condition of the state in Eq.~\eqref{eq:initialstate2}, $a^2+b^2+c^2=1$, we obtain $\sum_j E_j=\mathbb{I}_2$. Thus the $\{E_j\}$ satisfy the completeness relation. The probabilities of obtaining an outcome for an individual signal state is preserved under the mapping. When the sum of probabilities over different signal states or measurement outcomes is calculated, the probabilities are added. Hence, when considering only optimal measurements, the classical mutual information is invariant under the mapping. Thus the Holevo bound for the $2$-dimensional Hilbert space provides an upper bound for the accessible information in the $3$-dimensional Hilbert space. This proves $I^{\cal M}_{y,r}\le 1$.
\epf

\section{Proof of Proposition~\ref{prop1}}\label{app:proofp1}

\bpf
According to Protocol~\ref{ptl:NLAND}, the reduced density operator received by Bob has support in the Hilbert space spanned by orthonormal kets $\ket{0},\ket{1},\ket{2}$. In the following we consider the purification of such mixed state: assume Alice's input state to be a pure state on two qutrits, one of them being an ancillary qutrit belonging to Alice. This change from mixed states to pure states on a larger system would not decrease (and in fact it possibly increases) the information quantities in the left-hand side of the inequalities~\eqref{eq:info1}\eqref{eq:info2}\eqref{eq:info3}. Hence, if we can prove the inequalities for pure states on such enlarged system, we have proved the assertion.

We may assume the following form of Alice's input state (not in Schmidt form in general)
\bea\label{eq:initialstate}
a\ket{e_0}\ket{0}+b\ket{e_1}\ket{1}+c\ket{e_2}\ket{2},
\eea
where $a,b,c\ge 0$, and $a^2+b^2+c^2=1$, and $\ket{e_0},\ket{e_1},\ket{e_2}$ are unit vectors on the ancillary qutrit. The possible phases or signs in $a,b,c$ have been absorbed into $\ket{e_0},\ket{e_1},\ket{e_2}$. Since Bob's operation is only some phase gate on the second qutrit, Alice may apply a controlled unitary transform, with the second qutrit being the control, to make the transformed kets for $\ket{e_0},\ket{e_1},\ket{e_2}$ be orthogonal to each other. This unitary transform would have no effect on her ability (neither positively or adversely) in distinguishing the four returned states of Bob's. This explains why in the statement of the Proposition, we assume that Bob's received state $\sigma_A$ is the same for the two terms in the left-hand-side of each inequality, rather than assuming that Alice's initial pure states (including the part on her ancillary system) are the same. Using mixed states is also more natural in the sense that in Protocol~\ref{ptl:NLAND}, an honest Alice indeed uses one of several pure single-qutrit states, with the choice known to her, instead of using an entangled pure state.

From the last paragraph, for Alice to learn about $y$ and $r$, or their joint distribution, it is equivalent to assume that $\ket{e_0},\ket{e_1},\ket{e_2}$ are orthogonal to each other. Then for calculation of the information quantities in the inequalities~\eqref{eq:info1}\eqref{eq:info2}\eqref{eq:info3}, we could abbreviate Alice's ancillary qutrit (note that this is not a tracing-out operation but just a mathematical correspondence with a special purpose) and assume that the state initially sent by Alice is just a single qutrit state
\bea\label{eq:initialstate2}
a\ket{0}+b\ket{1}+c\ket{2},\quad \textrm{where}\,\, a,b,c\ge 0,
\eea
and the normalization of this state implies $a^2+b^2+c^2=1$. Note that this state is only for calculation of the information quantities but not the actual state used in the protocol. For example, if Alice is honest and chooses $x=0$ in Protocol~\ref{ptl:NLAND}, she uses an equal mixture of $\frac{1}{\sqrt{2}}(\ket{0}+\ket{2})$ and $\frac{1}{\sqrt{2}}(\ket{0}-\ket{2})$, and this can be replaced with a pure state $\frac{1}{2}(\ket{e_0}(\ket{0}+\ket{2})+\ket{e_1}(\ket{0}-\ket{2}))$. This can be written as $\frac{1}{2}((\ket{e_0}+\ket{e_1})\ket{0}+(\ket{e_0}-\ket{e_1})\ket{2})$. Ignoring the first qutrit (again, note that this is not a tracing-out operation), we have that the equivalent input state for calculation of the information quantities is $\frac{1}{\sqrt{2}}(\ket{0}+\ket{2})$.

In Protocol~\ref{ptl:NLAND}, the $y$ and $r$ are independent, and each may take the value $1$ with probability $\frac{1}{2}$. The four states after Bob's gate are as follows:
\bea
r=0,\, y=0:\quad && a\ket{0}+b\ket{1}+c\ket{2},\notag\\
r=0,\, y=1:\quad && a\ket{0}-b\ket{1}+c\ket{2},\notag\\
r=1,\, y=0:\quad && -a\ket{0}-b\ket{1}+c\ket{2},\notag\\
r=1,\, y=1:\quad && -a\ket{0}+b\ket{1}+c\ket{2}.\label{eq:newstate1}
\eea

In Eq.~\eqref{eq:info1}, the two measurements are the same. This means that Alice needs to use the same measurement $\cal M$ to learn information about $y$ and $r$. The four states in \eqref{eq:newstate1} are exactly the same as those in \eqref{eq:newstate0}. Therefore, Lemma~\ref{lemma1} implies that $I^{\cal M}_{y,r}\le 1$.

Also note that there are effectively no prior correlations between the two parties, so the locking of information \cite{DHL04} does not occur here. The above implies that the amount of information that Alice learns about the joint distribution of $y$ and $r$ is upper bounded by $1$ bit. The bits $y$ and $r$ are independent when Bob produces them, so the $y$ and $r$ are independent prior to Alice's measurement. Thus the inequality~\eqref{eq:info1} holds, where we have assumed that the two $\sigma_A$ implicit in the information quantities are the same in this equation (same below). The bits $y$ and $y\oplus r$ jointly determine $y$ and $r$, and vice versa, so the amount of information that Alice learns about the joint distribution of $y$ and $y\oplus r$ is upper bounded by $1$ bit. And since the bits $y$ and $y\oplus r$ are independent prior to Alice's measurement, we have that the inequality~\eqref{eq:info2} holds. The inequalities \eqref{eq:info1} and \eqref{eq:info2} together imply \eqref{eq:info3}. This completes the proof.
\epf

\section{Proof of Proposition~\ref{prop:probability}}\label{app:proofp2}

\bpf
Similar to the proof of Prop.~\ref{prop1}, we assume Alice's input state to be a two-qutrit pure state of the form \eqref{eq:initialstate} by introducing an ancillary qutrit, since this would not decrease the guessing probabilities as compared to using mixed states on one qutrit. In other words, we choose to prove a stronger assertion.

For the purpose of proving the assertion, it suffices to consider the input state as being on a $3$-dimensional Hilbert space, since Alice could do a unitary transform on the two-qutrit state of the form \eqref{eq:initialstate} preserving the guessing probabilities, to make it a linear combination of $\ket{00},\ket{11},\ket{22}$, and we may rewrite these basis states as $\ket{0},\ket{1},\ket{2}$. Hence the state is [the same as Eq.~\eqref{eq:initialstate2}]
\bea
a\ket{0}+b\ket{1}+c\ket{2},\label{eq:initialstate3}
\eea
where $a,b,c\ge 0$, and $a^2+b^2+c^2=1$. Then, after Bob's phase gate, the state becomes one of four states in Eq.~\eqref{eq:newstate1}. We can write out the density operators for $y=0$ and $y=1$ as (each after taking average over values of $r$)
\bea
\rho_{0}=(a\ket{0}+b\ket{1})(a\bra{0}+b\bra{1})+c^2\ketbra{2}{2},\label{eq:y0}\\
\rho_{1}=(a\ket{0}-b\ket{1})(a\bra{0}-b\bra{1})+c^2\ketbra{2}{2}.\label{eq:y1}
\eea
The trace distance [recall that it is defined as $D(\rho,\sigma)=\frac{1}{2}\tr\vert\rho-\sigma\vert$] of these two density operators is $2ab$, thus $P_y=\frac{1+2ab}{2}=\frac{1}{2}+ab$. Similarly, the average density operator for $r=0$ and $r=1$ (averaged over values of $y$) are as follows:
\bea
\tau_{0}=(a\ket{0}+c\ket{2})(a\bra{0}+c\bra{2})+b^2\ketbra{1}{1},\label{eq:r0}\\ \tau_{1}=(a\ket{0}-c\ket{1})(a\bra{0}-c\bra{1})+b^2\ketbra{2}{2}.\label{eq:r1}
\eea
The trace distance of these two density operators is $2ac$, thus $P_r=\frac{1+2ac}{2}=\frac{1}{2}+ac$. Hence,
\bea
(P_r-\frac{1}{2})^2+(P_y-\frac{1}{2})^2&=&a^2 b^2 + a^2 c^2 \notag\\
&=& a^2 (b^2+c^2) \notag\\
&=& a^2 (1-a^2) \notag\\
&\le& \frac{1}{4},\label{eq:guessingprob3}
\eea
And the equality is reached only when $a^2=\frac{1}{2}$, i.e. $a=\frac{1}{\sqrt{2}}$. The input states reaching the equality is given in the example in Appendix~\ref{app:example}.

For proving the second inequality in the assertion, we may similarly write out the density matrices for different values of $y\oplus r$, and obtain that the trace distance of these two density operators is $2bc$. Then remaining steps are  similar to those for the first inequality.
\epf

\section{Proof of Proposition~\ref{prop:Holevotradeoff}}\label{app:proofp3}

\bpf
(i) We denote $A=a^2, B=b^2, C=c^2$, then $A+B+C=1$, and $A,B,C\ge 0$. We may rewrite Eq.~\eqref{eq:chir} using $A$ and $C$ only:
\bea
\chi_r=-A\log_2 A-C\log_2 C+(A+C)\log_2(A+C),\label{eq:chir2}
\eea
and rewrite Eqs.~\eqref{eq:chiy} and \eqref{eq:chiyxorr} as
\bea
\chi_y=-A\log_2 A-B\log_2 B+(A+B)\log_2(A+B),\notag\\
\label{eq:chiy2}\\
\chi_{y\oplus r}=-B\log_2 B-C\log_2 C+(B+C)\log_2(B+C).\notag\\
\label{eq:chiyxorr2}
\eea
We first find a relation of $B$ and $\delta$.
Due to the concavity of the function $f(x)=-x\log_2 x$ for $x\in(0,1]$, when $B$ is fixed, i.e. when $A+C$ is fixed, the maximum of Eq.~\eqref{eq:chir2} is achieved when $A=C$. Thus
\bea
\chi_r&\le&-(1-B)\log_2\frac{(1-B)}{2}+(1-B)\log_2 (1-B)\notag\\
&=&1-B\label{eq:B}
\eea
Then
\bea
\delta=1-\chi_r\ge B.\label{eq:delta}
\eea
This implies $B\le\delta<0.5$.

The $\chi_y$ in ~\eqref{eq:chiy2} is a monotonic increasing function of $A$ when $0\le A\le 1-B$. Thus we take the value of $\chi_y$ when $A=1-B$ as an upper-bound estimate:
\bea
\chi_y\le -(1-B)\log_2 (1-B)-B\log_2 B=h(B).\label{eq:chir3}
\eea
The right-hand-side of \eqref{eq:chir3} is a monotonic increasing function of $B$ when $B<0.5$, and since $B\le\delta<0.5$, we obtain the inequality \eqref{eq:chitradeoff1}.
Similarly we can prove the inequality \eqref{eq:chitradeoff2} for $0\le\delta<0.5$.

(ii) The proof is completely similar to that of (i).
\epf

\section{Proof of Theorem~\ref{thm1}}\label{app:proofthm1}

\bpf
We first consider the case that Alice's operations are independent among different instances of Protocol~\ref{ptl:NLAND}, and then comment that the non-independent case still satisfies the extreme case of the inequalities for the first case, and discuss the effect of ``restarts'' on the security of Protocol~\ref{ptl:precompute1}. This gives rise to the security of Protocol~\ref{ptl:precompute1}.

Due to the freedom of measurement basis choice mentioned above, the Holevo bounds, which are upper bounds of the information quantities, are more relevant for proving the security of Protocol~\ref{ptl:precompute1}. Under the condition that Alice's operations are independent among the instances, we need only consider the Holevo bounds for a single instance of Protocol~\ref{ptl:NLAND}. Let $\chi_y$ be the Holevo quantity which is the upper bound for $I^{\cal M}_y$, see Eq.~\eqref{eq:defHolevo}. The definition of $\chi_y$ shows that it is conditioned on the uniform prior distribution for $y$. The quantities $\chi_r$ and $\chi_{y\oplus r}$ are defined similarly and are also conditioned on the uniform prior distribution for $y$. From Prop.~\ref{prop:Holevotradeoff}, the following inequality holds for small positive $\e<0.5$. [As in Prop.~\ref{prop:Holevotradeoff}, the function $h(\e)\equiv -(1-\e)\log_2 (1-\e)-\e\log_2 \e$.]
\bea\label{eq:holevo4}
\chi_y \le h(\e),\,\quad\,\mbox{for}\,\,\max(\chi_r,\chi_{y\oplus r})\ge 1-\e.
\eea

Alice may cheat in some instances of Protocol~\ref{ptl:NLAND}. We define the expected failure rate $\e$ as the expected number of wrong results in the untested instances of Protocol~\ref{ptl:NLAND} versus the total number of untested instances in a run of Protocol~\ref{ptl:precompute1}. It is sort of subjective for Bob to estimate $\e$ from the number of wrong results in the tested instances and the total number of tests in Protocol~\ref{ptl:precompute1}, since it depends on the \emph{a priori} knowledge about $\e$. It should be noted that for practical applications, in which two parties do want to perform some two-party secure computation, the prior probability distribution of $\e$ should not be too biased, i.e. it must contain a non-negligible part that corresponds to almost no failure, since otherwise no batch of one-time tables may pass Bob's test under reasonable criteria. There is also a practical way for Bob to estimate $\e$ based on the observed failure rate only: he can estimate $\e$ using $\e=\Theta(\frac{K_f+1}{K})$, where $\Theta(\cdot)$ represents the exact order equivalence, i.e. there are positive constants $c_a$ and $c_b$ such that $c_a \frac{K_f+1}{K}\le\e\le c_b\frac{K_f+1}{K}$, and the $K_f$ is the number of failed tested instances, and $K$ is the number of tested instances. (The $1$ appears here for avoiding the problem of vanishing $\e$ when $K_f=0$, which presents a problem for later analysis.)

Suppose that after some checking, Bob estimates that the expected failure rate is $\e$, then the following estimate holds for the remaining unchecked instances of Protocol~\ref{ptl:NLAND}, for the uniform distribution of $y$ and $r$ (the uniform distribution of $y$ can be imposed by Bob since he wants to make Alice's cheating be detected, and the $r$ has uniform distribution according to Protocol~\ref{ptl:NLAND}): $\max(\chi_r,\chi_{y\oplus r})\ge 1-c_1\e$, where $c_1$ is a positive constant, which arises because not all instances that passed checks are with Alice's honest behavior. Hence, $\chi_y\le h(c_1\e)$ according to Eq.~\eqref{eq:holevo4}. This shows that the expected amount of information about $y$ learnable by a cheating Alice in the remaining instances of Protocol~\ref{ptl:NLAND} is arbitrarily near zero for sufficiently small $\e$, even if she measures in different bases from those for the tested instances. The word ``expected'' means that even if $L\cdot h(c_1\e)<1$, where $L$ is the total number of one-time tables to be used for the main computation, Alice may sometimes learn about one or a few bits of Bob's input by chance, but on average, she learns not more than $L\cdot h(c_1\e)$ bits of information. Since $L$ is fixed and we can make $\e$ arbitrarily small by using more redundant checks, it is not necessary to state a condition such as ``on average'' in the assertion to be proved. Since the information about $y$ is linearly related to the information learnable by Alice in the later main computation stage (see the bipartite AND-gate computation method in Sec.~\ref{sec2}), this shows the security of Protocol~\ref{ptl:precompute1} in the case that Bob's operations are independent among instances of Protocol~\ref{ptl:NLAND}.

We give an estimate of the cost overhead due to checks, under the assumptions that Bob estimates $\e$ using $\e=\Theta(\frac{K_f+1}{K})$ and that the threshold for aborting is set to a constant number $c_0$ of failures. Conditioned on that the protocol has not aborted, the estimated $\e$ satisfies $\e=\Theta(\frac{c_0}{K})=\Theta(\frac{1}{K})$. From $\chi_y\le h(c_1\e)$, the $\e$ should satisfy $\e=o(\frac{1}{L})$ for the final one-time tables to contain less than one unsafe instances, where $L$ is the desired number of one-time tables. Thus $\frac{L}{K}=o(1)$, meaning that the number of tested instances $K$ is strictly larger than the order of $L$,  and this is the only requirement on $K$, thus $K$ being on the order of $O(L^{1.1})$ is sufficient. Also note that the number of remaining untested instances should be at least equal to $L$. Thus the total number of instances of Protocol~\ref{ptl:NLAND} is somewhat higher than $L$, but the overhead ratio can be quite small compared to $L$, say on the order of $O(L^{\nu})$ with $\nu$ being a small real number near $0$, say $\nu=0.1$.

In the following we consider the general case that Alice's operations are not necessarily independent among instances of Protocol~\ref{ptl:NLAND}. If Alice initially prepares some correlated quantum states among $M$ instances, the generalization of Eq.~\eqref{eq:holevo4} for the corresponding Holevo bounds should hold approximately near the extreme point $\e=0$, due to the uniform continuity of the Holevo bounds (as functions of the joint state received by Bob on multiple subsystems). Since Bob's variables $y$ and $r$ are independent among the instances, the generalizations of Eq.~\eqref{eq:holevo4} just mentioned have the same scaling near the extreme point $\e=0$ (as the number of instances of Protocol~\ref{ptl:NLAND} grows) as in the case that Alice's operations are independent. The last point can be seen from that Alice's states in other instances of Protocol~\ref{ptl:NLAND} serve as auxiliary systems in considering Holevo quantities of the form \eqref{eq:defHolevo}, and our proof of Prop~\ref{prop:Holevotradeoff} implicitly allowed auxiliary systems, because of the reduction from the case with auxiliary system to the case without such system in the proof of Prop.~\ref{prop1}. Thus the one-copy tradeoff curve of the Holevo quantities still holds, i.e. Eq.~\eqref{eq:holevo4} for one instance still holds with the same quantitative levels. This shows that the argument for the security for the case of independent operations of Alice can be extended to the general case. And the cost overhead estimate above also holds in this general case because the information upper-bound tradeoff relations are exactly similar.

Finally we consider the ``restarts'' of the protocol mentioned in the end of Protocol~\ref{ptl:precompute1}. Since Bob's inputs among different runs are independent, Alice has no way of using joint initial states or making joint measurements to take advantage of the possibility of restarts. Hence the probability that a cheating Alice would pass Bob's test adds up at most additively. And since practically there can only be a polynomial number of restarts, due to resource constraints, Bob can set appropriate thresholds in his checking to make the overall probability of cheater passing the tests upper bounded by any small positive constant.
\epf

\section{Proof of Proposition~\ref{prop:Holevo3}}\label{app:proofp4}

\bpf
In Protocol~\ref{ptl:NLAND}, Alice measures the received qutrit in the basis $\{\frac{1}{\sqrt{2}}(\ket{x}+\ket{2}),\frac{1}{\sqrt{2}}(\ket{x}-\ket{2}),\ket{1-x}\}$. The third measurement outcome is impossible in the ideal case, but actually, due to Bob's cheating, there is some possibility that the third measurement outcome occurs. For this outcome, it is natural to assume that Alice just guesses the outcome of $q$ randomly without bias, since she has no other side information in the current case of Protocol~\ref{ptl:precompute1b} (where Alice is honest in the initial stage) to give her any bias.

Suppose
\bea\label{eq:chisum3}
\frac{1}{2}\left[I^{M(\cU)}_{t \oplus r \vert{x=0},A'} + I^{M(\cU)}_{t \oplus r \vert{x=1},A'}\right]>1-\delta,
\eea
where small positive $\delta$ near $0$. Then since $I^{M(\cU)}_{t \oplus r \vert{x=0},A'}\le 1$ and $I^{M(\cU)}_{t \oplus r \vert{x=1},A'}\le 1$, we have
\bea\label{eq:chisum4}
I^{M(\cU)}_{t \oplus r \vert{x=x_0},A'}>1-2\delta,\,\,\mbox{for}\,\,x_0=0\,\,\mbox{and}\,\,1.
\eea
In the following we show that for each $x_0\in\{0,1\}$, there is a positive number $\eta$ such that
\bea\label{eq:prob0}
\bra{\psi_q}\rho_q\ket{\psi_q} \ge 1-2\delta,\,\,\mbox{for}\,\,q=0\,\,\mbox{and}\,\,1,
\eea
where $q=t\oplus r \vert_{x=x_0}$, and the $\ket{\psi_q}$ is the ideal state of the qutrit (with specific value of $q$) received by Alice, where ``ideal'' means Bob is honest; the $\rho_q$ is the actual state of the qutrit sent to Alice by Bob.
The proof of this fact is by contradiction. Suppose that
\bea\label{eq:prob1}
\bra{\psi_q}\rho_q\ket{\psi_q} < 1-2\delta
\eea
for some $q$ given that $x=x_0$. Then from the first paragraph of the proof, the probability that the state $\rho_q$ is recognized as $1-q$ by Alice is
\bea
P_{error}&=&\bra{\psi_{1-q}}\rho_q\ket{\psi_{1-q}}\notag\\
&&+\frac{1}{2}\left(1-\bra{\psi_q}\rho_q\ket{\psi_q}-\bra{\psi_{1-q}}\rho_q\ket{\psi_{1-q}}\right)\notag\\
&\ge& \frac{1}{2}\left(1-\bra{\psi_q}\rho_q\ket{\psi_q}\right) > \delta
\eea
Then, by assuming that $\rho_{1-q}$ is equal to the ideal state $\ketbra{\psi_{1-q}}{\psi_{1-q}}$, we obtain that the mutual information
\bea
I^{M(\cU)}_{t \oplus r \vert{x=x_0},A'}&<&\frac{1}{2}+\delta\log_2{\delta}-(\frac{1}{2}+\delta)\log_2{(\frac{1}{2}+\delta)}\notag\\
&=& \frac{1}{2}+\delta\log_2{\delta}-(\frac{1}{2}+\delta)[\log_2{(1+2\delta)-1]}\notag\\
&=& 1+\delta+\delta\log_2{\delta}-(\frac{1}{2}+\delta)\log_2{(1+2\delta)}\notag\\
&<& 1+\delta+\delta\log_2{\delta}\notag\\
&=& 1+\delta\log_2{2\delta}\notag\\
&<& 1-2\delta,\,\,\mbox{for}\,\,\delta\in(0,0.1),\label{eq:infodelta}
\eea
where the right-hand side of the first line is obtained by the mutual information when $P_{error}=\delta$. The joint probability distribution between the input and output for calculating such mutual information is $(\frac{1}{2},0,\delta,\frac{1}{2}-\delta)$.
The other possible choices of $\rho_{1-q}$ would only reduce the amount of mutual information. Thus we obtain a contradiction with the inequalities in \eqref{eq:chisum4}. Therefore, the assumption in ~\eqref{eq:prob1} is false, and the inequalities in \eqref{eq:prob0} are true.

In the following we prove that for $\delta$ near $0$, Bob's information about $x$ is limited, in the sense that $I^{M(\cU)}_{x,B'}=O(\delta^{1/4}\log\frac{1}{\delta})$.

In general we have to consider a possibly cheating Bob's unitary gates and measurements on the received state from Alice and his ancillary state, where some measurements may be prior to other gates. We always consider an equivalent circuit in which the measurements are deferred to the final steps. We can always insert a ``correct'' unitary gate $U(y,r)$ that an honest Bob should do, followed by $U^\dag(y,r)$, before the other unitary gates and measurements mentioned above. Note here that $y$ is an assumed bit and need not be some actual bit used in the protocol, since Bob may cheat about the value of $y$ he had used in the checking process later. So there is always a stage in the protocol when Bob does the correct gate $U(y,r)$, and at this point the qutrit state is one of the four pure states of the form \eqref{eq:newstate0} with $a=c=\frac{1}{\sqrt{2}}$, $b=0$ (when $x=0$) or $b=c=\frac{1}{\sqrt{2}}$, $a=0$ (when $x=1$). That is, when $x=0$, the state is one of the following:
\bea
t\oplus r=0:\quad && \ket{\tau_{00}}=\frac{1}{\sqrt{2}}(\ket{0}+\ket{2}),\notag\\
t\oplus r=1:\quad && \ket{\tau_{01}}=\frac{1}{\sqrt{2}}(\ket{0}-\ket{2}),\label{eq:newstatetau}
\eea
When $x=1$, the state is one of the following:
\bea
t\oplus y\oplus r=0:\quad && \ket{\tau_{10}}=\frac{1}{\sqrt{2}}(\ket{1}+\ket{2}),\notag\\
t\oplus y\oplus r=1:\quad && \ket{\tau_{11}}=\frac{1}{\sqrt{2}}(\ket{1}-\ket{2}).\label{eq:newstatetau2}
\eea

Consider the input states $\ket{\tau_{00}}$ and $\ket{\tau_{10}}$. The fidelity between these two states is
\bea
\vert\braket{\tau_{00}}{\tau_{10}}\vert=\frac{1}{2}.
\eea
Before Bob's final measurements, the circuit is unitary, so the inner product is preserved till this step. And Bob's later measurements, whether it is on $A'$ (before sending it to Alice) or $B'$ or at some step before splitting the system into $A'$ and $B'$, would not increase the information obtainable by Alice about $q$. Thus, to allow maximal information obtainable by Alice, we could just consider unitary circuits followed by local measurements on system $B'$. Denote the state on $A'B'$ before the measurements corresponding to $\ket{\tau_{00}}$ and $\ket{\tau_{10}}$ as $\ket{\xi_{00}}$ and $\ket{\xi_{10}}$, respectively. Since the inner product is preserved under unitaries, we have
\bea
\vert\braket{\xi_{00}}{\xi_{10}}\vert=\vert\braket{\tau_{00}}{\tau_{10}}\vert=\frac{1}{2}.
\eea
The reduced density operators on $A'$ for the states $\ket{\xi_{00}}$ and $\ket{\xi_{10}}$ are defined as follows:
\bea
{\rm Tr}_{B'}(\proj{\xi_{00}})&=&\rho_{00},\notag\\
{\rm Tr}_{B'}(\proj{\xi_{10}})&=&\rho_{10}.
\eea
According to the first part of the proof, which argues for that \eqref{eq:prob0} is true,
\bea\label{eq:prob2}
\bra{\tau_{q0}}\rho_{q0}\ket{\tau_{q0}} \ge 1-2\delta,\,\,\mbox{for}\,\,q=0\,\,\mbox{and}\,\,1.
\eea

Denote the reduced density matrices on $B'$ for the states $\ket{\xi_{00}}$ and $\ket{\xi_{10}}$ as follows
\bea
{\rm Tr}_{A'}(\proj{\xi_{00}})&=&\gamma_{00},\notag\\
{\rm Tr}_{A'}(\proj{\xi_{10}})&=&\gamma_{10}.
\eea
In the following we argue that $\gamma_{00}$ and $\gamma_{10}$ must be near each other for $\vert\braket{\xi_{00}}{\xi_{10}}\vert=\frac{1}{2}$ to hold. Suppose the Schmidt decompositions of $\ket{\xi_{00}}$ and $\ket{\xi_{10}}$ are
\bea\label{eq:schmidt}
\ket{\xi_{00}}&=&\sum_j f_j\ket{a_j}\ox\ket{b_j},\notag\\
\ket{\xi_{10}}&=&\sum_k g_k\ket{a'_k}\ox\ket{b'_k},
\eea
where $f_j,g_k$ are real positive numbers satisfying $\sum_j f_j^2=\sum_k g_k^2=1$, and $\{\ket{a_j}\}$ (and $\{\ket{a'_k}\}$) is a set of orthogonal normalized states on $\cH_{A'}$, and $\{\ket{b_j}\}$ (and $\{\ket{b'_k}\}$) is a set of orthogonal normalized states on $\cH_{B'}$.

From Eqs.~\eqref{eq:prob2} and \eqref{eq:schmidt},
\bea\label{eq:schmidt2}
\sum_j f_j^2 \vert\braket{a_j}{\tau_{00}}\vert^2\ge 1-2\delta,\notag\\
\sum_k g_k^2 \vert\braket{a'_j}{\tau_{10}}\vert^2\ge 1-2\delta.
\eea
there is at least one ket among $\{\ket{a_j}\}$ that satisfies $\vert\braket{a_j}{\tau_{00}}\vert\ge \sqrt{1-2\delta}$.
Since the kets $\ket{a_j}$ are orthogonal, and $\delta<0.1$, there can only be one ket $\ket{a_j}$ that satisfies this requirement. We denote this special $j$ as $\hat j$. Similarly, we denote the $k$ such that $\vert\braket{a'_k}{\tau_{10}}\vert\ge \sqrt{1-2\delta}$ as $\hat k$. These can be expressed as
\bea\label{eq:schmidt3}
\vert\braket{a_{\hat j}}{\tau_{00}}\vert&\ge& \sqrt{1-2\delta},\notag\\
\vert\braket{a'_{\hat k}}{\tau_{10}}\vert&\ge& \sqrt{1-2\delta}.
\eea

In the following we obtain an upper bound for $\vert\braket{a_j}{\tau_{00}}\vert$, where $j\ne{\hat j}$.
Using the formula $\cos(a-b)=\cos(a)\cos(b)+\sin(a)\sin(b), \forall a,b\in\mathbb{R}$, for $j\ne{\hat j}$ we have
\bea\label{eq:otherfj}
\vert\braket{a_j}{\tau_{00}}\vert & \le & \vert\braket{a_j}{a_{\hat j}}\vert\cdot\vert\braket{a_{\hat j}}{\tau_{00}}\vert\notag\\
&&+\sqrt{1-\vert\braket{a_j}{a_{\hat j}}\vert^2}\sqrt{1-\vert\braket{a_{\hat j}}{\tau_{00}}\vert^2}\notag\\
&=&0+\sqrt{1-\vert\braket{a_{\hat j}}{\tau_{00}}\vert^2}\notag\\
&\le&\sqrt{2\delta},\quad \forall j\ne{\hat j}.
\eea
We will now obtain a lower bound for $f_{\hat j}$. From the first inequality in Eq.~\eqref{eq:schmidt2},
\bea
f_{\hat j}^2 \vert\braket{a_{\hat j}}{\tau_{00}}\vert^2 &\ge&  \sqrt{1-2\delta}-\sum_{j\ne{\hat j}} f_j^2 \vert\braket{a_j}{\tau_{00}}\vert^2\notag\\
&\ge& \sqrt{1-2\delta}-\sum_{j\ne{\hat j}}f_j^2 (2\delta)\notag\\
&=& \sqrt{1-2\delta}-(1-f_{\hat j}^2) 2\delta,
\eea
Thus
\bea\label{eq:fj}
f_{\hat j}^2 &\ge& (\sqrt{1-2\delta}-2\delta)/(\vert\braket{a_{\hat j}}{\tau_{00}}\vert^2-2\delta)\notag\\
&\ge& \frac{\sqrt{1-2\delta}-2\delta}{1-2\delta}\notag\\
&\ge& 1-2\delta,\,\,\mbox{for}\,\,\delta\in\,(0,0.1),
\eea
implying that $f_{\hat j}\ge\sqrt{1-2\delta}$. Similarly, we have
\bea\label{eq:gk}
g_{\hat k}^2\ge 1-2\delta,
\eea
implying that $g_{\hat k}\ge\sqrt{1-2\delta}$.

Consider approximations to $\ket{\xi_{00}}$ and $\ket{\xi_{10}}$ as follows:
\bea
\ket{\xi'_{00}}&=&\ket{\tau_{00}}\ox\ket{b_{\hat j}}=\ket{\tau_{00}}\ox\ket{B_{00}},\notag\\
\ket{\xi'_{10}}&=&\ket{\tau_{10}}\ox\ket{b'_{\hat k}}=\ket{\tau_{10}}\ox\ket{B_{10}},
\eea
where $\ket{B_{00}}\equiv\ket{b_{\hat j}}$, $\ket{B_{10}}\equiv\ket{b'_{\hat k}}$.

Then we have
\bea\label{eq:fidelity}
\vert\braket{\xi_{00}}{\xi'_{00}}\vert&=& f_{\hat j}\vert\braket{a_{\hat j}}{\tau_{00}}\vert\ge 1-2\delta,\notag\\
\vert\braket{\xi_{10}}{\xi'_{10}}\vert&=& g_{\hat k}\vert\braket{a'_{\hat k}}{\tau_{10}}\vert\ge 1-2\delta.
\eea

Noting that $\ket{\xi'_{00}}$ and $\ket{\xi'_{10}}$ are both product states, we have
\bea
\vert\braket{\xi'_{00}}{\xi'_{10}}\vert&=&\vert\braket{\tau_{00}}{\tau_{10}}\vert\cdot \vert\braket{B_{00}}{B_{10}}\vert\notag\\
&=&\frac{1}{2}\vert\braket{B_{00}}{B_{10}}\vert.\label{eq:compare2}
\eea
Recall that $\vert\braket{\xi_{00}}{\xi_{10}}\vert=\frac{1}{2}$, and using the formula $\cos(a+b)=\cos(a)\cos(b)-\sin(a)\sin(b),\,\forall a,b\in\mathbb{R}$, we have
\bea
\vert\braket{\xi_{10}}{\xi'_{00}}\vert&\ge&\vert\braket{\xi_{00}}{\xi_{10}}\vert \cdot \vert\braket{\xi_{00}}{\xi'_{00}}\vert\notag\\
&&-\sqrt{1-\vert\braket{\xi_{00}}{\xi_{10}}\vert^2}\sqrt{1-\vert\braket{\xi_{00}}{\xi'_{00}}\vert^2}\notag\\
&\ge& \frac{1}{2}(1-2\delta)-\frac{\sqrt{3}}{2}\sqrt{4\delta-4\delta^2}\notag\\
&\ge& \frac{1}{2}(1-2\delta)-\sqrt{3\delta}\ge \frac{1}{2}-3\sqrt{\delta}\label{eq:compare1a}
\eea
Then
\bea
\vert\braket{\xi'_{00}}{\xi'_{10}}\vert&\ge& \vert\braket{\xi_{10}}{\xi'_{00}}\vert \cdot \vert\braket{\xi_{10}}{\xi'_{10}}\vert\notag\\
&&-\sqrt{1-\vert\braket{\xi_{10}}{\xi'_{00}}\vert^2}\sqrt{1-\vert\braket{\xi_{10}}{\xi'_{10}}\vert^2}\notag\\
&\ge&(\frac{1}{2}-3\sqrt{\delta})(1-2\delta)\notag\\
&&-\sqrt{1-(\frac{1}{2}-3\sqrt{\delta})^2}\sqrt{1-(1-2\delta)^2}\notag\\
&\ge& \frac{1}{2}-3\sqrt{\delta}-\delta-\sqrt{\frac{3}{4}+3\sqrt{\delta}}\sqrt{4\delta}\notag\\
&\ge& \frac{1}{2}-3\sqrt{\delta}-\delta-\sqrt{3}(1+2\sqrt{\delta})\sqrt{\delta}\notag\\
&=& \frac{1}{2}-(3+\sqrt{3})\sqrt{\delta}-(1+2\sqrt{3})\delta\notag\\
&\ge& \frac{1}{2}-7\sqrt{\delta},\,\,\mbox{for}\,\,\delta\in(0,0.1)\label{eq:compare1}
\eea
From Eqs.~\eqref{eq:compare1} and \eqref{eq:compare2}, we have
\bea\label{eq:innerproduct}
\vert\braket{B_{00}}{B_{10}}\vert\ge 1-14\sqrt{\delta}.
\eea

Let $D(\rho,\sigma)\equiv \frac{1}{2}{\rm Tr}\vert\rho-\sigma\vert$ be the trace distance. According to \cite{NC00},
\bea\label{eq:trdistance}
D(\rho,\sigma)\le \sqrt{1-F(\rho,\sigma)^2}.
\eea
And the trace distance satisfies the triangle inequality, therefore
\bea
D(\gamma_{00},\gamma_{10})&\le& D(\gamma_{00},\ket{B_{00}})+D(\ket{B_{00}},\ket{B_{10}})\notag\\
&&+D(\gamma_{10},\ket{B_{10}})\notag\\
&\le& 2[2\cdot 2\delta]+\sqrt{1-(1-14\sqrt{\delta})^2}\notag\\
&\le& \sqrt{28}\delta^{1/4}+8\delta=O(\delta^{1/4}),
\eea
where the first term in the second line is from Eqs.~\eqref{eq:fj}\eqref{eq:gk}. To see this, note that the contribution to $D(\gamma_{00},\ket{B_{00}})$ from those $\ket{b_j}$ with $j\ne {\hat j}$ in the pure-state decomposition of $\gamma_{00}=\sum_j f_j^2 \ketbra{b_j}{b_j}$ is not greater than $2\delta$, due to that $\sum_{j\ne {\hat j}} f_j^2=1-f_{\hat j}^2\le 2\delta$, and the other contribution is from that $f_{\hat j}^2$ is at most $2\delta$ from $1$, in the term $f_{\hat j}^2 \ketbra{b_{\hat j}}{b_{\hat j}}$. For pairs of states among $\ket{\xi_{00}}$, $\ket{\xi_{10}}$, $\ket{\xi_{01}}$ and $\ket{\xi_{11}}$ with different values of $x$, we get similar relations between their reduced density operators on $\cH_{B'}$. The information obtainable by Bob about $x$ is related to the distinguishability of the reduced density operators on $\cH_{B'}$ for the states $\ket{\xi_{00}}$ and $\ket{\xi_{01}}$ versus those for the states $\ket{\xi_{10}}$ and $\ket{\xi_{11}}$. Therefore, the accessible information obtainable by Bob about $x$ is upper bounded by $O[h(\delta^{1/4})]$, where $h(\e)\equiv -(1-\e)\log_2 (1-\e)-\e\log_2 \e$. For $\delta$ sufficiently small, this amount of information can be expressed as $O(\delta^{1/4}\log\frac{1}{\delta})$. This completes the proof.
\epf

\section{Proof of Theorem~\ref{thm2}}\label{app:proofthm2}

\bpf
First, we consider the case that Bob's operations are independent among instances of Protocol~\ref{ptl:NLAND}. In such case we need only consider the information tradeoff inequalities for a single instance of Protocol~\ref{ptl:NLAND}. From Prop.~\ref{prop:Holevo3}, when
\bea\label{eq:chisum3ar}
\frac{1}{2}\left[I^{M(\cU)}_{t \oplus r \vert{x=0},A'} + I^{M(\cU)}_{t \oplus r \vert{x=1},A'}\right]>1-\delta,
\eea
where $\delta$ is a small positive constant near $0$, then
\bea\label{eq:tradeoff7r}
I^{M(\cU)}_{x,B'}=O(\delta^{1/4}\log\frac{1}{\delta})
\eea
for sufficiently small $\delta$.

Bob may cheat in some instances of Protocol~\ref{ptl:NLAND}. We define the expected failure rate $\e$ as the expected number of wrong results in the untested instances of Protocol~\ref{ptl:NLAND} versus the total number of untested instances in a run of Protocol~\ref{ptl:precompute1b}. It is sort of subjective for Alice to estimate $\e$ from the number of wrong results in the tested instances and the total number of tests in Protocol~\ref{ptl:precompute1b}, since it depends on the \emph{a priori} knowledge about $\e$. It should be noted that for practical applications, in which two parties do want to perform some two-party secure computation, the prior probability distribution of $\e$ should not be too biased, i.e. it must contain a non-negligible part that corresponds to almost no failure, since otherwise no batch of one-time tables may pass Alice's test under reasonable criteria. There is also a practical way for Alice to estimate $\e$ based on the observed failure rate only: she can estimate $\e$ using $\e=\Theta(\frac{K_f+1}{K})$, where the notations are the same as in the proof of Theorem~\ref{thm1}. In particular, the $K_f$ is the number of failed tested instances, and $K$ is the number of tested instances.

Suppose that after some checking, Alice estimates that the expected failure rate is $\e$. Since in Protocol~\ref{ptl:NLAND} Alice finally learns $t\oplus (x\cdot y)\oplus r$ provided she knows $x$, the condition about the $\e$ can be expressed as $\frac{1}{2}\left[I^{M(\cU)}_{t \oplus r \vert{x=0},A'} + I^{M(\cU)}_{t \oplus r \vert{x=1},A'}\right]\ge 1-c_2\e$ for the remaining untested instances of Protocol~\ref{ptl:NLAND}, where $c_2$ is a positive constant, which arises because not all instances that passed checks are with Bob's honest behavior. Then, the result of Prop.~\ref{prop:Holevo3} implies $I^{\cU}_{x,B'}\le c_3 (c_2\e)^{1/4}\log\frac{1}{\e}$ for those instances. This shows that the expected amount of information about $x$ learnable by a cheating Bob in the remaining instances of Protocol~\ref{ptl:NLAND} is arbitrarily near zero for sufficiently small $\e$, even if he performs different unitaries $\cU$ (followed by arbitrary measurements on his part of the output) from those for the tested instances. Since the information about $x$ is linearly related to the information learnable by Bob in the later main computation stage (see the bipartite AND-gate computation method in Sec.~\ref{sec2}), this shows the security of Protocol~\ref{ptl:precompute1b} in the case that Bob's operations are independent among instances of Protocol~\ref{ptl:NLAND}.

We give an estimate of the cost overhead due to checks, under the assumptions that Alice estimates $\e$ using $\e=\Theta(\frac{K_f+1}{K})$, and that the threshold for aborting is set to a constant number of failures. Conditioned on that the protocol has not aborted, the estimated $\e$ satisfies $\e=\Theta(\frac{c_0}{K})=\Theta(\frac{1}{K})$. Since $I^{\cU}_{x,B'}\le c_3 (c_2\e)^{1/4}\log\frac{1}{\e}$, the $\e$ should satisfy $\e=o(\frac{1}{L^4})$ for the final one-time tables to contain less than one unsafe instances, where $L$ is the desired number of one-time tables. Thus $\frac{L^4}{K}=o(1)$, meaning that the number of tested instances $K$ is strictly larger than the order of $L^4$, and this is the only requirement on $K$, thus $K$ being on the order of $O(L^{4.1})$ is sufficient. The overhead ratio is on the order of $O(L^{3+\nu})$ with $\nu$ being a small real number near $0$, say $\nu=0.1$.

In the following we consider the general case that Bob's operations (including possible measurements after the unitary $\cU$) are not independent among instances of Protocol~\ref{ptl:NLAND}. Note that Alice's input bits $x$ and the bits $t$ are independent among the copies. The generalizations of the information quantities in the inequalities \eqref{eq:chisum3ar} and \eqref{eq:tradeoff7r} to the multi-copy case can be easily defined, and from the proof of Prop.~\ref{prop:Holevo3} it can be seen that a tradeoff of multi-copy information quantities should satisfy the similar relation, just with the right-hand-side of the inequality \eqref{eq:chisum3ar} multiplied by $M$, the number of copies of Protocol~\ref{ptl:NLAND}. The systems from the other instances of Protocol~\ref{ptl:NLAND} serve as auxiliary systems for one instance. This shows that the security holds for the general case that Bob's operations are not independent among instances of Protocol~\ref{ptl:NLAND}. And the cost overhead estimate above also holds in this general case because the information tradeoff relations are exactly similar.

For the ``restarts'' of the protocol, the argument is exactly similar to that in the proof of Theorem~\ref{thm1}, but it is stated for the case that both parties perform some checking, instead of just one party. We abbreviate it here.
\epf

\section{Examples for Protocol~\ref{ptl:NLAND}}\label{app:example}

\textbf{Example 1.} In the following we show a continuous family of Alice's input states reaching the equality in Eq.~\eqref{eq:info1}, as well as the equality in Eq.~\eqref{eq:guessingprob}.
The states are one-qutrit states
\begin{eqnarray}
\frac{1}{\sqrt{2}}(\ket{0}+\cos\alpha\ket{1}+\sin\alpha\ket{2}),\label{eq:beststate}
\end{eqnarray}
where $\alpha\in [0,\frac{\pi}{2}]$ is a real parameter. The qutrit is sent to Bob.  After Bob does his operations on the received qutrit and sends it back to Alice, an optimal measurement of Alice to recover information about the joint distribution of $y$ and $r$ is a POVM measurement with $4$ POVM elements, and they are of the form
\begin{eqnarray}
&\frac{1}{2}(\cos\alpha\ket{0}+\ket{1})(\cos\alpha\bra{0}+\bra{1}),&\notag\\
&\frac{1}{2}(\cos\alpha\ket{0}-\ket{1})(\cos\alpha\bra{0}-\bra{1}),&\notag\\
&\frac{1}{2}(\sin\alpha\ket{0}+\ket{2})(\sin\alpha\bra{0}+\bra{2}),&\notag\\
&\frac{1}{2}(\sin\alpha\ket{0}-\ket{2})(\sin\alpha\bra{0}-\bra{2}).&\label{eq:povm}
\end{eqnarray}
The four POVM elements above sum up to the identity operator on the $3$-dimensional Hilbert space. Under such choice of input state and measurement, it can be calculated that $I^{\cal M}_y=\cos^2\alpha$, and $I^{\cal M}_r=\sin^2\alpha$. The sum of these quantities is $1$, and since $y$ and $r$ are independent, such measurement gives $1$ bit of classical mutual information between the measurement outcomes and the joint distribution of $y$ and $r$. Therefore, the equality in \eqref{eq:info1} and the equality $I^{\cal M}_{y,r}=1$ both hold for the input state in \eqref{eq:beststate} and the POVM measurement in \eqref{eq:povm} (Note that these equations may not simultaneously hold for other states, such as those in Example 3). The reason why this family of input states satisfy the equality in Eq.~\eqref{eq:guessingprob} is that the condition $a^2=\frac{1}{2}$ in the proof of Prop.~\ref{prop:probability} is satisfied.

\textbf{Example 2.} Similarly, there is a family of input states on two qutrits satisfying the equality in Eq.~\eqref{eq:info1} and the equality in Eq.~\eqref{eq:guessingprob}. The states are
\begin{eqnarray}
\frac{1}{\sqrt{2}}(\ket{00}+\cos\alpha\ket{11}+\sin\alpha\ket{22}),\label{eq:beststate2}
\end{eqnarray}
where $\alpha\in [0,\frac{\pi}{2}]$ is a real parameter, and the first qutrit is withheld by Alice, and the second qutrit is sent to Bob. After Bob does his operations on the received qutrit and send it back to Alice, an optimal measurement of Alice to recover information about the joint distribution of $y$ and $r$ is a POVM measurement with $5$ POVM elements, with four of the POVM elements of the form
\begin{eqnarray}
&\frac{1}{2}(\cos\alpha\ket{00}+\ket{11})(\cos\alpha\bra{00}+\bra{11}),&\notag\\
&\frac{1}{2}(\cos\alpha\ket{00}-\ket{11})(\cos\alpha\bra{00}-\bra{11}),&\notag\\
&\frac{1}{2}(\sin\alpha\ket{00}+\ket{22})(\sin\alpha\bra{00}+\bra{22}),&\notag\\
&\frac{1}{2}(\sin\alpha\ket{00}-\ket{22})(\sin\alpha\bra{00}-\bra{22}).&\label{eq:povm2}
\end{eqnarray}
The four listed POVM elements sum up to the identity operator on the $3$-dimensional subspace spanned by $\{\ket{00},\ket{11},\ket{22}\}$, and the remaining POVM element may be chosen as the projector onto the orthogonal subspace spanned by the remaining $6$ computational-basis states. Under such choice of input state and measurement, it can be calculated that $I^{\cal M}_y=\cos^2\alpha$, and $I^{\cal M}_r=\sin^2\alpha$. For the similar reason as in the previous example, the equality in \eqref{eq:info1} and the equality $I^{\cal M}_{y,r}=1$ both hold for the input state in \eqref{eq:beststate2} and the POVM measurement in \eqref{eq:povm2}. Similarly, this family of input states satisfy the equality in Eq.~\eqref{eq:guessingprob}.\\

\textbf{Example 3.}
The following suspected information sum value [the left-hand-side of \eqref{eq:info1}] for a generic input state of the form \eqref{eq:initialstate2} is first found by numerical calculation, and analytical construction for the measurement that reaches the suspected information sum value is given, but we do not have a proof that the constructed measurement is optimal (except for the case that the information sum is already $1$). For input state of the form \eqref{eq:initialstate2}, and noting that $a,b,c\ge 0$, we rewrite the state as
\bea
a\ket{0}+\sqrt{1-a^2}\cos\theta\ket{1}+\sqrt{1-a^2}\sin\theta\ket{2},\notag\\
\quad \textrm{where}\,\, a\ge 0,\,\,\theta\in [0,\frac{\pi}{2}],\label{eq:initialstate4}
\eea
then the left-hand-side of \eqref{eq:info1} is at least
\bea
(a + b')^2 \log_2 (a + b') + (a - b')^2 \log_2 \vert a - b'\vert, \label{eq:infomutual1}
\eea
where $b'\equiv \sqrt{1-a^2}$. The measurement that reaches this amount of information sum is exactly the same as the POVM measurement shown in Eq.~\eqref{eq:povm}.
When $b=b'$, i.e. when $c=0$, this measurement becomes a projective measurement in the basis $\{\frac{1}{\sqrt{2}}(\ket{0}+\ket{1}),\frac{1}{\sqrt{2}}(\ket{0}-\ket{1}),\ket{2}\}$. When $a=\frac{1}{\sqrt{2}}$, the expression in \eqref{eq:infomutual1}, i.e. the left-hand-side of \eqref{eq:info1}, is equal to $1$, which agrees with the result in Example 1. We also calculated $I^{\cal M}_{y,r}$ where $\cal M$ is some POVM measurement implemented by projective measurement on enlarged Hilbert space ($9$-dimensional), and the maximum found numerically is generically greater than the expression in \eqref{eq:infomutual1} when $a<\frac{1}{\sqrt{2}}$ and $bc>0$, and sometimes it may even be close to $1$. But when $a>\frac{1}{\sqrt{2}}$, the maximum numerical values of $I^{\cal M}_{y,r}$ and $I^{\cal M}_y+I^{\cal M}_r$ are quite near. We also calculated $I^{\cal M}_y+I^{\cal M'}_r$ for different POVM measurements $\cal M$ and $\cal M'$, and the maximum sum found numerically is usually greater than the expression in \eqref{eq:infomutual1} when $a\ne \frac{1}{\sqrt{2}}$ and $bc>0$. The similar result holds for input state of the form \eqref{eq:initialstate} (with the form of the optimal measurement changed accordingly).

\section{Numerical results}\label{app:num}

In this appendix, we present some numerical results.
The inequalities about classical mutual information in Prop.~\ref{prop1} is verified by numerical calculation of random (cheating) states and some special classes of (cheating) states. Since the calculation of mutual information involves maximizing over possible measurements, and we used ancilla of limited dimensions in the measurement, the results are only indicative, and do not prove the inequalities in Prop.~\ref{prop1}. But since the Holevo bound is an upper bound of accessible information, and is easier to calculate since it does not involve maximizing over measurements, the numerical results below about the Holevo bounds are more convincing and provide checks against the results about mutual information. We have found and numerically verified that some continuous families of states satisfy the equality in Eq.~\eqref{eq:info1} (and hence Eq.~\eqref{eq:info3}), and every point in the tradeoff curve of the two terms is reachable, see Eqs.~\eqref{eq:beststate} and \eqref{eq:beststate2}. Similar continuous family of states which satisfy the equality in Eq.~\eqref{eq:info2} can be written out by symmetry. It is interesting to note that some states which satisfy the equalities require no ancilla, but note that POVM measurements are needed. If there is no ancilla initially, and only projective measurements on $3$ dimensions are used, then the equality may be reached at the end of the tradeoff curve of the two terms, but near the middle of the curve, we can only find values of the sum being slightly less than $0.9$ bit at most for some tested classes of input states; we have not attempted exhausting all possible input states for such point.

We have also found that, if we remove the requirement that the two measurements in the left-hand side of Eq.~\eqref{eq:info1} (and similarly, Eqs.~\eqref{eq:info2} and \eqref{eq:info3}) be the same, then it is possible to get a sum larger than $1$ on the left-hand side. The numerical value obtained, when not using ancilla and using two possibly different \emph{projective} measurements, is already larger than $1.2$ bits for the input state $\frac{1}{\sqrt{2}}\ket{0}+\frac{1}{2}\ket{1}+\frac{1}{2}\ket{2}$. But of course, the obtained sums are not greater than the sum of Holevo bounds shown below.

The Figure~\ref{fig:holevo} shows the tradeoff relation for the Holevo quantities, $\max(\chi_r,\chi_{y\oplus r})$ and $\chi_y$, arising from Alice's cheating states in Protocol~\ref{ptl:NLAND}. The calculation allows for Alice's possible cheating by using an initial entangled state of two qutrits, where one of the qutrits is sent to Bob. The number of sampled states is $50$ million. The curve shows the maximum of the vertical coordinates among the samples in the same small range of length $0.01$ (called a ``bin'') over the horizontal axis. The ideal curve should be symmetric with respect to the two axes. The imperfections in the left part of the curve are believed to be due to insufficient number of samples, and the inherent asymmetry in the taking the maximum of the vertical coordinate in each bin. The maximum sum of the values of the two coordinates is about $1.38848$ bits, which is approximately achieved when the value of the two coordinates are about equal. Numerics suggest that near the ends of the tradeoff curve, one coordinate approaches 1 (bit) while the other coordinate approaches 0, confirming Eq.~\eqref{eq:holevo4}.

Assuming that the maximum sum of Holevo quantities (the maximum sum of two coordinates) in the figure is achieved when the two coordinates are equal (in particular we assume $\chi_y=\chi_r$), we can obtain an analytical expression for the maximum and the corresponding coordinates, as well as the corresponding parameters of the input state. From Eqs.~\eqref{eq:chiy} and \eqref{eq:chir}, we have
\bea
&&\chi_y+\chi_r\notag\\
&&=-a^2\log_2 a^2-b^2\log_2 b^2+(1-c^2)\log_2(1-c^2)\notag\\
&&-a^2\log_2 a^2-c^2\log_2 c^2+(1-b^2)\log_2(1-b^2),\label{eq:chisum}
\eea
then from $b^2=c^2$ which follows from the assumption $\chi_y=\chi_r$ and the apparent fact that $a>0$ for achieving the maximum sum, and using $a^2=1-b^2-c^2$, we have
\bea
&&\chi_y+\chi_r\notag\\
&&=-2a^2\log_2 a^2-2b^2\log_2 b^2+2(1-b^2)\log_2(1-b^2),\notag\\
&&=-2(1-2b^2)\log_2 (1-2b^2)\notag\\
&&\quad-2b^2\log_2 b^2+2(1-b^2)\log_2(1-b^2)\label{eq:chisum2}
\eea
By taking the derivative of the expression above with respect to the variable $b^2$, we have
that the maximum is achieved when $b^2=c^2=\frac{5-\sqrt{5}}{10}$. The corresponding $a^2=\frac{\sqrt{5}}{5}$.
The maximum of $\chi_y+\chi_r$ (the maximum sum of the two coordinates in the figure, after comparing the values of $\chi_r$ and $\chi_{y\oplus r}$) is $\log_2 (3+\sqrt{5})-1\approx 1.3884838$ bits.

\begin{figure}[ht]
\centering
\includegraphics[scale=0.13]{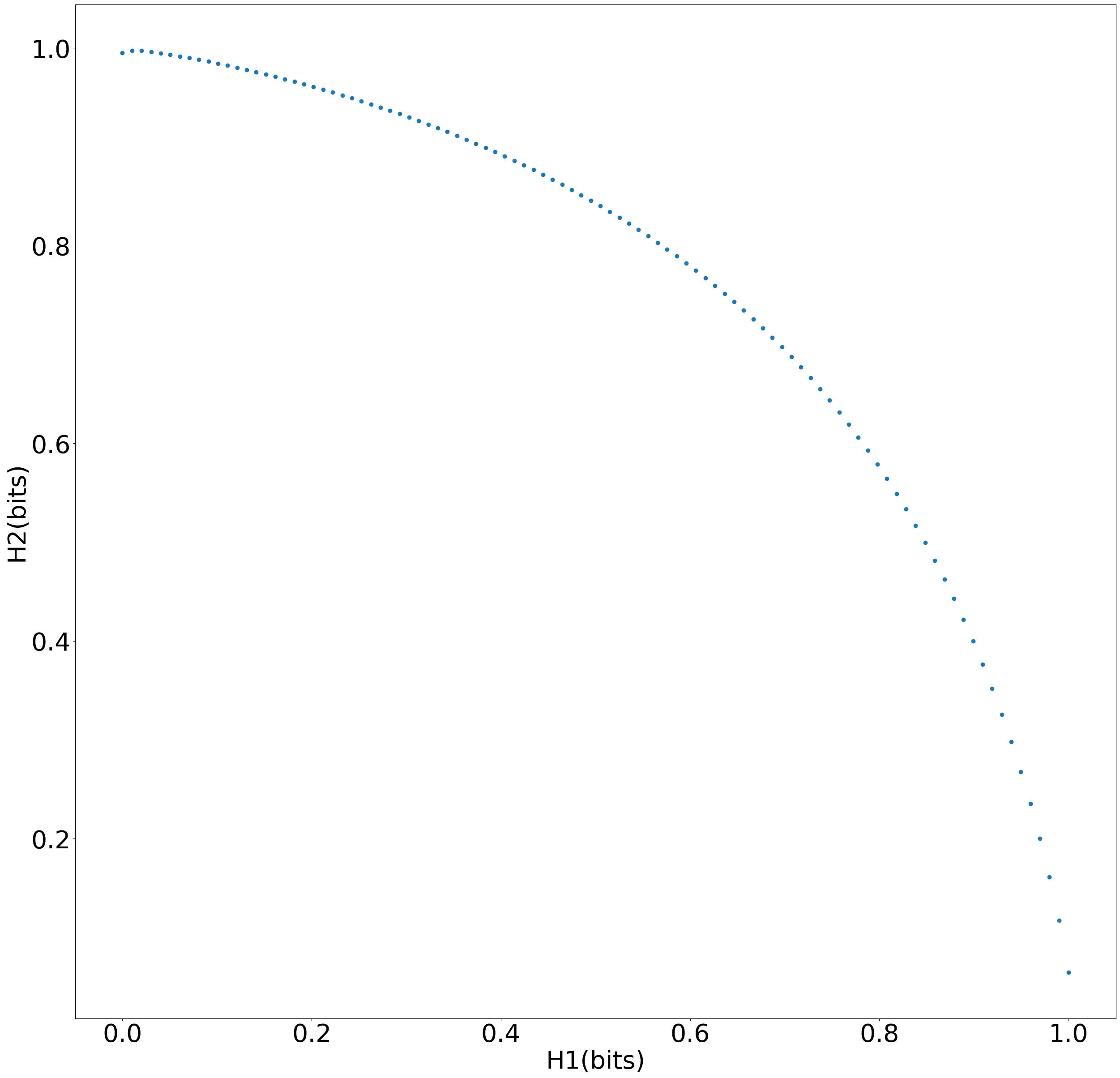}
\caption{An illustration of the tradeoff relations of the Holevo quantities by numerical calculations. Horizontal axis (H1): $\max(\chi_r,\chi_{y\oplus r})$; vertical axis (H2): $\chi_y$. These Holevo bounds are calculated from random pure input states on two qutrits, where one of the qutrits is withheld by Alice. In Bob's view, he receives a qutrit in some mixed state, performs a gate on it and send it back to Alice. The curve shows the maximum of the vertical coordinates among the samples in the same small range of length $0.01$ over the horizontal axis.}
\label{fig:holevo}
\end{figure}

\end{appendix}

\end{document}